\documentclass[fleqn,usenatbib]{mnras}

\usepackage{newtxtext,newtxmath}

\usepackage[T1]{fontenc}

\DeclareRobustCommand{\VAN}[3]{#2}
\let\VANthebibliography\thebibliography
\def\thebibliography{\DeclareRobustCommand{\VAN}[3]{##3}\VANthebibliography}

\usepackage{soul}	
\usepackage{xcolor}

\usepackage{graphicx}	
\usepackage{amsmath}	

\title[DSA and multimessenger flux from UFOs]{Diffusive shock acceleration at EeV and associated multimessenger flux from ultra-fast outflows driven by Active Galactic Nuclei}

\author[E. Peretti et al.]{Enrico Peretti$^{1}$\thanks{E-mail: enrico.peretti.science@gmail.com},
Alessandra Lamastra$^{2}$, Francesco Gabriele Saturni$^{2,3}$, 
\newauthor
Markus Ahlers$^{1}$,
Pasquale Blasi$^{4,5}$, 
Giovanni Morlino$^{6}$ and Pierre Cristofari$^{7}$
\\
\\
$^{1}$ Niels Bohr International Academy, Niels Bohr Institute,University of Copenhagen, Blegdamsvej 17, DK-2100 Copenhagen, Denmark \\
$^{2}$ INAF -- Osservatorio Astronomico di Roma, Via Frascati 33, I-00078 Monte Porzio Catone (RM), Italy\\
$^{3}$ ASI -- Space Science Data Center, Via del Politecnico snc, I-00133 Roma, Italy\\
$^{4}$ Gran Sasso Science Institute, Via F. Crispi 7, 67100, L’Aquila, Italy\\
$^{5}$ INFN/Laboratori Nazionali del Gran Sasso, Via G. Acitelli 22, 67100, Assergi (AQ), Italy\\
$^{6}$ INAF, Osservatorio Astrofisico di Arcetri, L.go E. Fermi 5, I-50125 Firenze, Italy\\
$^{7}$ Observatoire de Paris, PSL Research University, LUTH, 5 Place J. Janssen, 92195 Meudon, France}

\date{Accepted XXX. Received YYY; in original form ZZZ}

\pubyear{2015}

\begin{document}
\label{firstpage}
\pagerange{\pageref{firstpage}--\pageref{lastpage}}
\maketitle

\begin{abstract}
Active galactic nuclei (AGN) can launch and sustain powerful winds featuring mildly relativistic velocity and wide opening angle. Such winds, known as ultra-fast outflows (UFOs), can develop a bubble structure characterized by a forward shock expanding in the host galaxy and a wind termination shock separating the fast cool wind from the hot shocked wind. In this work we explore whether diffusive shock acceleration can take place efficiently at the wind termination shock of UFOs. We calculate the spectrum of accelerated particles and find that protons can be energized up to the EeV range promoting UFOs to promising candidates for accelerating ultra-high energy cosmic rays (UHECRs). We also compute the associated gamma-ray and neutrino fluxes and compare them with available data in the literature. We observe that high-energy (HE) neutrinos are efficiently produced up to hundreds of PeV while the associated gamma rays could be efficiently absorbed beyond a few tens of GeV by the optical-ultraviolet AGN photon field. By assuming a typical source density of non-jetted AGN we expect that UFOs could play a dominant role as diffuse sources of UHECRs and HE neutrinos. We finally apply our model to the recently observed NGC1068 and we find out that under specific parametric conditions an obscured UFO could provide a sizeable contribution to the observed gamma-ray flux while only contributing up to $\sim 10\%$ to the associated neutrino flux.
\end{abstract}




\section{Introduction}

Fast wide angle winds are one of the most intriguing feedback mechanisms of Active Galactic Nuclei (AGN) \citep[][]{Silk_Rees_1998}. Their impact on the host galaxies has long been considered to affect the dynamical evolution of the interstellar medium (ISM) and act as a regulator of star formation \citep[][]{Crenshaw_2003}. The discovery of blue-shifted Fe K absorption lines in X-ray spectra of AGN \citep[see e.g.][]{Chartas_2002} brought compelling evidence of mildly relativistic velocities typically ranging from $0.1 \, c$ to $ 0.3 \, c$ (where $c$ is the speed of light) in such winds, which thereafter have been often referred to as ultra-fast outflows (UFOs). 
Mildly relativistic flows were already known to be present in the vicinity of AGN from broad absorption lines observed in their spectra \citep[][]{Weymann_1991}. The discovery of UFOs allowed us to understand that high kinetic luminosity \citep[$\Dot{E} \simeq 10^{41}-10^{45} \rm erg \, s^{-1}$][]{Tombesi2015,Fiore_AGN} was also possible in fast AGN-driven winds with wide opening angles.
Recently UFOs have been systematically detected in both radio-quiet and radio-loud AGN \citep[see e.g.][]{Markovitz2006,Braito2007,Cappi2009,Tombesi_2010,Tombesi_2010_2,Tombesi2015}. The number of observations keeps increasing with time, despite of the observational challenges, also thanks to high resolution grating spectra in the soft X-ray \citep[see e.g.][]{Pounds_2203} and the discovery of ultraviolet (UV) lines in addition to those already known in the X-ray \citep[][]{UV_UFO1,UV_UFO_2}. The search for a launching mechanism of the UFOs has not found a definitive answer yet, although there is a general agreement to ascribe this phenomenon to the accretion activity of the super massive black hole (SMBH) \citep[see][for an updated review]{Laha_2021}. 

The dynamics of the wind and the associated feedback on the host galaxy depend on whether the outflow conserves energy or momentum \citep[][]{King_Pounds_2015}. In particular, if the wind plasma does not cool efficiently when it shocks with the surrounding medium, the system is energy-conserving and the momentum flux is boosted while the wind sweeps up the external matter. 
In the opposite scenario, namely if the wind radiates most of its thermal energy, it evolves conserving the momentum. In spite of the strong radiation fields that could strongly affect the cooling of electrons, \citet{Faucher-Giguere-2012} showed that two-temperature plasma effects are likely to slow down radiative losses for protons thereby favoring an energy-conserving dynamics.

A recent Fermi-LAT analysis \citep[][]{UFO-Fermi-LAT+Caprioli} showed that UFOs are a new class of gamma-ray emitters. 
In this analysis the average gamma-ray emission from a sample of 11 nearby (z<0.1) radio-quiet AGN with an UFO is derived by adopting a stacking analysis. The average best-fit gamma-ray spectral slope is measured to be $2.1 \pm 0.3$, and the gamma-ray luminosity is found to scale with the AGN bolometric luminosity.

AGN-driven outflows, similar to stellar winds \citep[see e.g.][]{Weaver77,Koo-McKee,Koo-McKee2}, are expected to develop a structure characterized by an inner wind termination shock (hereafter \textit{wind shock}), a contact discontinuity and an outer forward shock. The forward shock has been proposed as a plausible site for particle acceleration \citep[see e.g.][]{Lamastra_2016,Wang_Loeb_Gamma,Lamastra_2019,UFO-Fermi-LAT+Caprioli} where ideal conditions for efficient production of gamma rays and high-energy (HE) neutrinos are expected \citep[see also][for a recent study on molecular outflows]{McDaniel_Ajello2023}. 
\citet{Wang_Loeb_UHECR} highlighted also the possibility that, in somewhat extreme conditions, a fast AGN-driven wind could have the energy budget to accelerate CRs up to the ultra-high-energy (UHE) range.
The associated cumulative contribution of AGN-driven winds to the diffuse gamma-ray and neutrino flux has been also explored \citep[see e.g.][]{Wang_Loeb_Gamma,Wang_Loeb_neutrinos,Lamastra_2017,Liu_2018}. In addition, the amplitude of the recently observed spectrum of the diffuse neutrino flux \citep[][]{IceCube2020_LAST}, in light of the constraints imposed by the diffuse gamma-ray flux observed by Fermi-LAT \citep[][]{Fermi-LAT-diffused}, suggests that there could be a class of sources at least partially opaque to gamma rays \citep[see e.g.][]{Murase_hidden}. 

Indeed, a search for time-integrated point-like neutrino sources \citep[][]{IceCube_pointlike} highlighted an excess in the direction of the Seyfert galaxy NGC1068. 
The most intriguing aspect of the emission of such galaxy is the lack of gamma rays in the TeV band \citep[][]{MAGIC-UL-NGC1068}, where the neutrino flux is observed. The natural implication of an highly opaque cosmic particle accelerator triggered several studies on the multimessenger implications of HE particles populating the innermost region of AGN such as disks and accretion flows \citep[see e.g.][]{Kimura_RIAF,Gutierrez_riaf}, combined emission from successful and failed AGN winds \citep[][]{Susumu_Inoue_NGC1068} and AGN corona \citep[see e.g.][]{Murase_AGN_cores,Inoue2020,Kheirandish_2021,Foteini_NGC1068,Murase2022}. Interestingly, we recently witnessed a growth in the statistical significance of the signal from NGC1068 \citep[][]{IceCube-NGC-last}. 

Even though there is an increasing evidence for HE particles populating the innermost regions of active galaxies, our understanding of the acceleration mechanisms at play in such environments is still incomplete. 
The mechanisms capable of energizing HE particles in the vicinity of SMBH is indeed a partially unexplored field we aim to assess in this work together with its multimessenger consequences.

Therefore, we develop a model for particle acceleration exploring the diffusive shock acceleration (DSA) mechanism at the wind shock of UFOs. 
At this shock, unique conditions for acceleration of protons at energies as high as $\sim 10^{18}$ eV can be found. 
In addition, the medium property could make such sources bright in HE neutrinos while being partially opaque to gamma rays beyond $10-10^2$ GeV.
In particular, the optical thickness to gamma-rays depends on the specific parametric conditions.
We discuss the role of UFOs as UHE cosmic ray (UHECR) sources in light of the spectral behavior of the particle flux escaping the AGN wind bubble. 
In particular, we show that at the highest energies spectral features harder than $E^{-2}$ could appear in the spectrum of escaping particles due to the interplay of diffusion-advection and energy losses. 
This result is of great interest in light of phenomenological studies aiming at modeling the spectrum and mass composition observed by the Pierre Auger Observatory which suggest UHECRs to be characterized by hard spectra \citep[see e.g.,][and references therein]{Unger_2015}. 
Moreover, if UFOs were common in galaxies this would produce important implications for their diffuse multimessenger emission. We provide here an order of magnitude estimate of the potential role of UFOs for a diffuse flux of HE neutrinos and CR at the \textit{ankle}. 
Finally, we explore whether an UFO could play a role in the neutrino emission observed in NGC1068 discussing possible realizations of such a system which better agree with observed data.

The manuscript is organized as follows: in \S~\ref{Sec: 2-Wind-and-model} we describe the model for the wind bubble and the associated particle acceleration and transport formalism; in \S~\ref{Sec: 3-Results} we discuss our results in terms of spectra of accelerated and escaping particles, we perform a parameter space scan focusing on the maximum energy. In \S~\ref{Sec: Gamma and Nu} we discuss the multimessenger implications in terms of HE photons and neutrinos and in \S~\ref{Sec: Discussion-NGC1068}, we specialize our model by applying it to the case of NGC1068 and discuss possible model improvements and alternative scenarios. 
We draw our conclusions in \S~\ref{Sec: 5-Conclusions}.

\section{Model for particle acceleration and multimessenger emission in UFO}
\label{Sec: 2-Wind-and-model}

\begin{figure}\centering
	\includegraphics[width= 0.9\columnwidth]{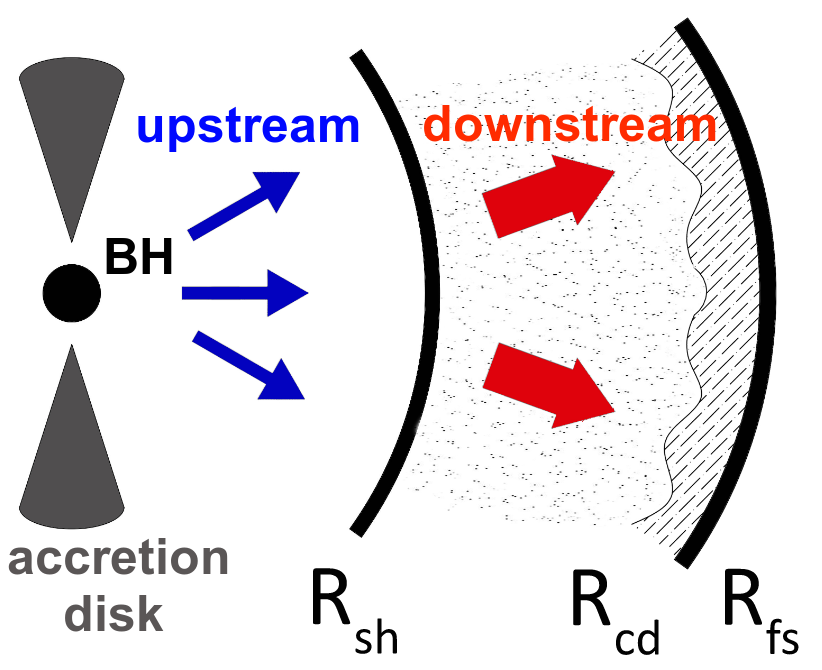} 
    \caption{Structure of the wind bubble. The SMBH (BH) responsible for the wind launching is located on the left of the sketch. The blue (red) arrows correspond to the cool (shocked) wind of the upstream (downstream) region. The wind shock ($R_{\rm sh}$) separates these two regions. The SAM is located between the contact discontinuity ($R_{\rm cd}$) and the forward shock ($R_{\rm fs}$) which bounds the system (credit: I. Peretti).}
    \label{fig:Image_1}
\end{figure}

The fast wind launched and sustained by AGN expands with large opening angle and mildly relativistic velocity. At such speed the outflow is supersonic, therefore it drives a forward shock expanding in the external medium, while a contact discontinuity separates the shocked wind (SW) material from the shocked ambient medium (SAM). 
The impact of the wind on the external medium creates a shock inside the wind material, the wind shock, which is trailing behind the contact discontinuity and it is oriented towards the central engine. 
The outflow, characterized by these three discontinuities, features a bubble structure as sketched in Figure~\ref{fig:Image_1}. 

The environment surrounding active SMBHs where UFOs expand is extremely complex: from the innermost parts of AGN one can find the accretion disc, the broad line region (BLR) with highly dense clouds of density up to $n_c \lesssim 10^{10} \, \rm cm^{-3}$ \citep[see e.g.][]{Ricci_2022}, a dusty torus, the narrow line region with clouds of typical density of about $10^4 \, \rm cm^{-3}$ and often a larger circum-nuclear disk \citep[see e.g.][]{Urry-Padovani1995}. 
For the purposes of modelling the dynamics of the UFO we assume a uniform circum-nuclear medium (CNM) of effective density $n_0$.
A plausible range of values for $n_0$ can be derived from the column density $N_H$, based on the work by \citet{Ricci_2017}: considering $N_H \lesssim 10^{25} \, \rm cm^{-2}$ as a typical range for the column density of AGN and assuming $10^6-10^9 \, \rm M_{\odot}$ as a plausible mass range for SMBHs, we obtain $1.5-63$ pc as a typical radius for the sphere of influence of the black hole ($GM_{\rm SMBH}/\sigma_*^2$) and an upper limit to the external medium density of $n_0 \lesssim 5 \cdot 10^4 - 10^6\, \rm cm^{-3}$. Here, we adopted the relation between black hole mass and stellar velocity dispersion, $M_{\rm SMBH}/10^9 {\rm M_{\odot}}\simeq 0.309 \times (\sigma_{*}/ 200 {\rm km \, s^{-1}})^{4.38}$\citep[][]{Kormendi2013}.

In a uniform medium of density $n_0$ the wind expands with constant velocity up to the radius at which the swept-up mass roughly balances the whole mass of the outflow. After the swept-up mass becomes dynamically relevant, the outflow starts decelerating. During the deceleration phase the forward shock $R_{\rm fs}$ and the wind shock $R_{\rm sh}$ evolve self-similarly according to different scaling laws: $R_{\rm fs} \sim t^{3/5}$ and $R_{\rm sh} \sim t^{2/5}$ \citep[see also][for detailed discussions and Appendix~\ref{Appendix: Self-similar radii evo} for additional details]{Weaver77,Koo-McKee,Koo-McKee2}. 
Since the wind shock decelerates faster than the forward shock, the hot bubble, namely the spherical shell between the wind shock and the contact discontinuity, grows with time while remaining approximately adiabatic. 
In this context, the wind bubble evolution can be considered as energy-conserving \citep[see e.g.,][]{Faucher-Giguere-2012}. On the other hand, radiative losses in the SAM can be efficient \citep[see e.g.][]{Nims_2015}, so that the whole swept-up mass is eventually compressed into a relatively thin layer between the contact discontinuity, $R_{\rm cd}$, and $R_{\rm fs}$. 
While the wind bubble slows down its expansion, the relative velocity between the plasma and the wind shock remains high, namely the shock stays strong. Therefore,
we refer to the innermost region of free expanding wind and to the shocked wind respectively as \textit{upstream} and \textit{downstream}. 
Different from the wind shock, the Mach number of the forward shock strongly depends on the temperature and conditions of the surrounding medium. 
Therefore, it is not guaranteed that the forward shock is strong for a sufficient amount of time. 

We assume a spherically symmetric geometry for the outflow and, since the wind launching region has a negligible size compared to the whole bubble, we also assume a constant upstream velocity $u_1$. 
The shocked wind is adiabatic, therefore the velocity profile reads: $u_2(r) = u_2 (R_{\rm sh}/r)^2$, where $u_2 = u_1/4$. In agreement with observations of UFOs, we limit our investigation to a plasma velocity lower than $\sim0.3 c$. 
We thus neglect relativistic effects due to the mildly relativistic motion of the plasma. 
The gas density in the upstream region scales as $n_1(r) = \Dot{M}/[4 \pi r^2 u_1 m_p]$, while in the downstream region it is constant and equal to $n_2 = 4 n_1(R_{\rm sh})$. 
The gas density between $R_{\rm cd}$ and $R_{\rm fs}$ depends on the amount of matter accumulated during the outflow evolution, $n_{\rm SAM} = n_0 R_{\rm fs}^3/(R_{\rm fs}^3-R_{\rm cd}^3)$, where we assume $R_{\rm cd} \simeq 0.9 \, R_{\rm fs}$ \citep{Sharma2014}. 
Under this assumption one obtains $n_{\rm SAM} \approx 4 n_0$ as a typical gas density in the SAM layer.
We postulate a turbulent nature for the magnetic field and we estimate its amplitude in the upstream region under the assumption that a fraction $\epsilon_B \lesssim 10 \%$ of the ram pressure is converted into magnetic field energy density, namely $U_B(r) = \epsilon_B  m_p n_1(r) u_1^2$. 
At the wind shock we assume that the magnetic field gets compressed by a factor $\sqrt{11}$, typical of strong shocks, and remains constant throughout the whole downstream region. 
We adopt the quasi-linear theory of diffusion, $D(r,p)= v(p) r_L^{2-\delta}(r,p) l_c^{\delta-1}/3$, where $v$ is the particle velocity, $r_L$ is the Larmor radius, $\delta$ is the slope of the turbulence power spectrum and $l_c$ is the coherence length of the magnetic field that we assume to be comparable in size with the launching radius of the wind. 
In addition, we account for the small angle scattering regime of diffusion, namely $D\propto r_L^2$, which takes place when $r_L>l_c$ \citep[see e.g.][]{Subedi2017,Dundovic2020}. 

\begin{figure}\centering
	\includegraphics[width= 1.0\columnwidth]{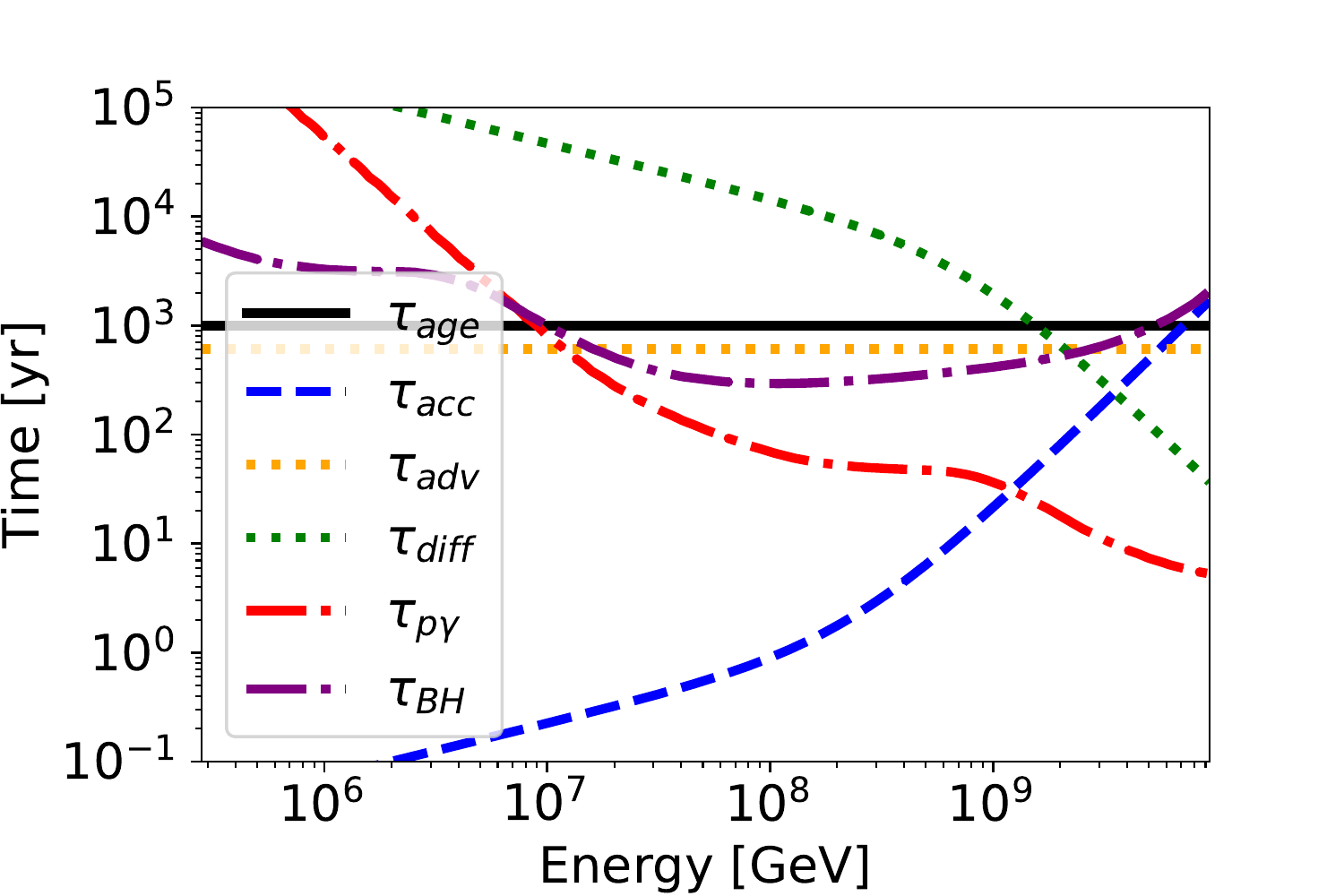} 
    \caption{Typical timescales regulating the transport of HE particles compared with the age of the system (thick black line). The blue dashed line represents the acceleration timescale, while energy losses via photomeson and Bethe-Heitler pair-production are represented respectively by red and magenta dot-dashed lines. 
    Advective and diffusive escape are represented by orange and green dotted lines.}
    \label{fig:Image_timescales}
\end{figure}

The average lifetime of AGN is inferred to be $\lesssim$10$^7$ yr \citep{yu02}. 
During this time, the AGN is expected to show multiple episodes of activity with duty cycles of $\lesssim$10$^5$ yr duration \citep{sch15}. 
This suggests that, even though their typical age is not known at present, UFOs could have a lifetime ranging from hundred years up to several thousands of years. 
In this work, we explore UFOs under the assumption that they can be powered for a sufficient amount of time so as to allow them to reach the deceleration phase. 
Therefore, $\rm O(10^3 \, \rm yr)$ is a conservative assumption while we comment in \S~\ref{SubSec: Param-Scan} the impact of assuming a different lifetime up to values comparable with the AGN duty cycle.
In this context, the dynamical evolution of the system becomes soon slower than all relevant timescales involving HE particles. Hence the process of particle acceleration and transport can be treated as stationary (see Figure~\ref{fig:Image_timescales} in Sec.~\S~\ref{Sec: 3-Results} where the typical timescales for HE particles in a prototype UFO are discussed).
We assume a spherically symmetric transport where particles are injected via DSA at the wind shock whereas, once they reach the forward shock location, they freely escape the wind bubble. 
The transport equation reads:
\begin{equation}
    \label{Eq: Transport}
    r^2 u \partial_r f = \partial_r[r^2 D \partial_r f] + \frac{p}{3} \partial_p f \, \partial_r[r^2 u] + r^2 [Q - \lambda f]\,, 
\end{equation}
where $u=u(r)$ is the wind velocity profile, $D=D(r,p)$ is the diffusion coefficient, $Q=Q(r,p)$ is the injection term and $\lambda=\lambda(r,p)$ is the loss rate accounting for pp and p$\gamma$ interactions (see Appendix~\ref{Appendix: Transport-solution} for additional details). As boundary conditions, consistently with the spherical symmetry, we assume a null net flux at the center of the system, $u f - D \partial_r f|_{r=0}=0$, while, as mentioned earlier, we regard the forward shock as a free escape boundary, $f(R_{\rm fs},p)=0$.
The injection term reads:
\begin{equation}
    \label{Eq: Injection}
    Q(r,p) = Q_0(p) \delta[r-R_{\rm sh}] = \frac{\eta_{\rm CR} u_1 n_1}{4 \pi p^2} \delta[p-p_{\rm inj}] \delta[r-R_{\rm sh}],
\end{equation}
where $p_{\rm inj} = 1 \, \rm GeV / c$ is the injection momentum of particles (picked up from the plasma) that enter the DSA process and $\eta_{\rm CR}$ is the efficiency factor, normalized such that the CR pressure at the shock is a small fraction ($\lesssim 10\%$) of the plasma ram pressure.

\begin{table}
\caption{Parameters of the benchmark UFO and of the three alternative scenario considered for the multimessenger emission.}
\label{table:parameters-0}
\centering             
\linespread{1.15}\selectfont
\begin{tabular}{c|c|c|c|c}
\hline
\hline   
Parameter & benchmark & A & B & C   \\
\hline
$u_1/c $ &  $0.2$ & - & - & -  \\
$ \Dot{M} [\rm M_{\odot} \, yr^{-1}]$ &  $10^{-1}$ & - & - & - \\
$\xi_{\rm CR}$ & $0.05$ & $0.087$ & $0.1$ & $0.12$ \\
$\epsilon_{\rm B} $ & $0.05$ & - & - & -  \\
$l_c [ \rm pc] $ & $10^{-2}$ & - & - & -  \\
$\delta$ & $3/2$ & - & - & - \\
$L_X  [\rm erg \, s^{-1}]$ & $10^{44}$ & - & - & -  \\
$n_{0} [\rm cm^{-3}]$ & $10^4$ & $2 \cdot 10^3$ & $5 \cdot 10^2$ & $2 \cdot 10^2$ \\
$t_{\rm age} [\rm yr] $ & $10^3$ & $3 \cdot 10^3$ & $10^4$ & $2 \cdot 10^4$  \\
\hline                   
\end{tabular}
\linespread{1.0}\selectfont
\end{table}

We solve Equation~\eqref{Eq: Transport} following the same procedure developed in \citet{Morlino_starcluster} and \citet{Peretti_wind} with the modification that, in the present work, energy losses in the downstream region are also accounted for due to their possible observational relevance.
In fact the relative small distance from an active SMBH makes the environment potentially hostile for HE particles where energy losses could affect the acceleration and limit the escape.
\\
The details of the calculation are reported in Appendix~\ref{Appendix: Transport-solution} while the general form of the solution at the wind shock reads:
\begin{equation}
\label{Eq: Solution-SHock-Text}
    f_{\rm sh}(p) = C p^{-s} {\rm e}^{-\Gamma_{\rm cut}(p)},
\end{equation}
where $C$ is a constant (see Appendix~\ref{Appendix: Transport-solution}), $s$ is the spectral index ($s=4$ in strong shocks) and $\Gamma_{\rm cut}$ is a HE cut-off function ($\Gamma_{\rm cut}=\Gamma_l+\Gamma_e$ as detailed in Appendix~\ref{Appendix: Transport-solution}) which increases with momentum.

As the background photon field, we use the spectral energy distribution (SED) model shown in the top panel of Figure~\ref{fig:Image_BKG_fields} provided in Appendix~\ref{Appendix: Self-similar radii evo}. 
Such a SED is characterized by the big blue bump and an X-ray power-law component as described in \citet{Marconi_2004_SED}. Such a field is assumed to decrease with the second power of distance from the central engine. Consistently with the AGN SED we also account for the far infrared (FIR) component produced by a dusty torus \citep{Mullaney11_torus}.
While the $r^{-2}$ profile is adequate for the photon field produced by the accretion disk and AGN corona (optical and X-rays), we point out that the FIR radiation field produced by the dusty torus is probably characterized by a more complex spatial structure \citep{Blazejowski2000}, possibly with a uniform behavior up to a radius comparable with the typical size of the torus ($\approx 1 \rm \, pc$). 
In Sec.~\ref{SubSec: Param-Scan} we discuss the implications and possible limits of such approximation.

Gamma rays and neutrinos from pp and p$\gamma$ interactions are computed following \citet{Kelner_pp} and \citet{Kelner_pg}, respectively. 
The energy losses due to Bethe-Heitler pair-production are taken into account as energy loss mechanism taking place at a rate $t^{-1}_{\rm BH}$ \citep[see][]{Gao2012-BH}.
Gamma-gamma absorption on the AGN SED including the torus is also accounted for by adopting the cross section appropriate for the case of an isotropic photon field \citep[see][]{Aharonian_book2004}. Finally, the associated flux-density of HE protons escaping the system is computed self-consistently from the solution to the transport equation as $j_{\rm esc}(p) = - D_2 \partial_r f |_{r=R_{fs}}$. 

The calculations illustrated here have been carried out in the context of the thin shell approximation, in which the SW is perfectly separated from the SAM. This implies that the cold gas acting as a target for pp interactions (see below) is all located close to the $R_{\rm fs}$. It is worth pointing out that the spatial distribution of the gas in the SW might be affected by instabilities and mixing that may lead to a more pervasive, though clumpy structure of the gas, which in turn might affect the spatial and spectral properties of the secondary emission. These effects will be investigated in a future dedicated work.

\section{Results}
\label{Sec: 3-Results}

In order to present our model and discuss its physical implications we assume a set of typical parameters, hereafter referred to as our \textit{benchmark} scenario. 
The parameters defining our benchmark scenario, summarized in Table~\ref{table:parameters-0}, have been chosen according to the following criteria: we assume $u_1=0.2 \, c$ as the average value for the terminal wind speed of UFOs; $\Dot{M}= 10^{-1} \, \rm M_{\odot} \, yr^{-1}$ has been chosen such that the total kinetic power $\Dot{E} = \Dot{M} u_1^2/2$ matches about $\sim 3\%$ of the total AGN bolometric luminosity \citep[][]{Fiore_AGN} characterized by an X-ray luminosity $L_X=10^{44} \, \rm erg \, s^{-1}$;
$l_c = 10^{-2} \, \rm pc$ is compatible with the launching radius of the wind as predicted by accretion disk wind models \citep[see e.g.][]{Murray1995-Launching}; the age of the system $t_{\rm age} = 10^3 \, \rm yr$ has been chosen in order to assure stationary conditions, which are not guaranteed for much younger systems (the resulting shock radii at such an age are $R_{\rm sh}\approx 0.8 \, \rm pc$ and $R_{\rm fs}\approx 3 \, \rm pc$); $\epsilon_B = 0.05$ guarantees a minor dynamical impact of the turbulent magnetic field; $\delta = 3/2$ is motivated by an MHD-like (Kraichnan) turbulence cascade.  
In agreement with the upper limits presented in Section~\ref{Sec: 2-Wind-and-model}, we assume $n_{0} = 10^4 \, \rm cm^{-3}$ as a typical value for the external ambient medium  that can be found also in the core of luminous infrared galaxies \citep[see e.g.][]{Downes_gal_density,Faucher-Giguere-2012} and we discuss in \S~\ref{SubSec: Param-Scan} the impact of adopting densities up to $n_0=10^6 \, \rm cm^{-3}$. 
Higher densities could be reached if the bulk of the column density were concentrated within a smaller volume consistent with the size of the BLR, $R_{\rm BLR}\lesssim 0.1 \, \rm pc$ \citep{Bentz}. However, the investigation of such a scenario goes beyond the limits of applicability of our model because, in such dense and compact environment, the system would be in calorimetric conditions. Nevertheless, we discuss possible implications of ultra-dense environments in \S~\ref{Subs: 1068 disc}.

Even though we study particle acceleration and transport by solving the stationary transport equation, Equation~\eqref{Eq: Transport}, a more direct understanding of the physics property of the solution can be obtained by analyzing the typical timescales of the different competing processes.
Figure~\ref{fig:Image_timescales} illustrates the typical timescales for HE particles as computed at $R_{\rm sh}$ for the benchmark scenario. 
Here, the age of the system ($\tau_{\rm age}$, thick black line) is compared with the following timescales: acceleration ($\tau_{\rm acc} \approx s D_1/u_1^2$, blue dashed line), diffusion ($\tau_{\rm diff}=(R_{\rm esc}-R_{\rm sh})^2/D_2$, green dotted line the diffusion), advection ($\tau_{\rm adv}=(R_{\rm esc}-R_{\rm sh})/<u_2>$, orange dotted), the $p \gamma$ photomeson ($\tau_{p \gamma}$, red dot-dashed line) and Bethe-Heitler pair-production ($\tau_{\rm BH}$, magenta dot-dashed).
Inelastic pp collisions were also taken into account, however at the wind shock the target density is of the order of $20 \, \rm cm^{-3}$, so that the associated timescale exceed $10^5 \, \rm yr$, the upper limit of the plot. 
Therefore, pp interactions are irrelevant for the acceleration. 
On the other hand, this does not exclude them as relevant loss mechanism in the SAM where the density is orders of magnitude higher.
From the interplay between the different timescales it is possible to observe: 
1) $\tau_{\rm acc} \ll \tau_{\rm age}$ and the minimum between losses and escape is also smaller than the age supporting the stationary assumption; 
2) $\tau_{\rm acc}$ as well as $\tau_{\rm diff}$ feature a break between $10^2$ PeV and 1 EeV due to the transition in the diffusion coefficient from the QLT behavior ($r_{\rm L}<l_{\rm c}$) to the small angle scattering regime ($r_{\rm L}>l_{\rm c})$; 
3) energy losses via $p \gamma$ photomeson production play a dominant role at the highest energies and are expected to set the maximum energy; 
4) $\tau_{p \gamma}$ and $\tau_{\rm BH}$ increase with the second power of the distance moving outward from $R_{\rm sh}$ to $R_{\rm fs}$, therefore the transport in the downstream region is characterized by a close competition between energy losses and escape;
5) the Bethe-Heitler pair production does not play a dominant role since at low energy the transport is advection-dominated while at the highest energies is regulated by the photomeson production.

\begin{figure}
\includegraphics[width=\columnwidth,viewport = 0 0 410 265,clip=true]{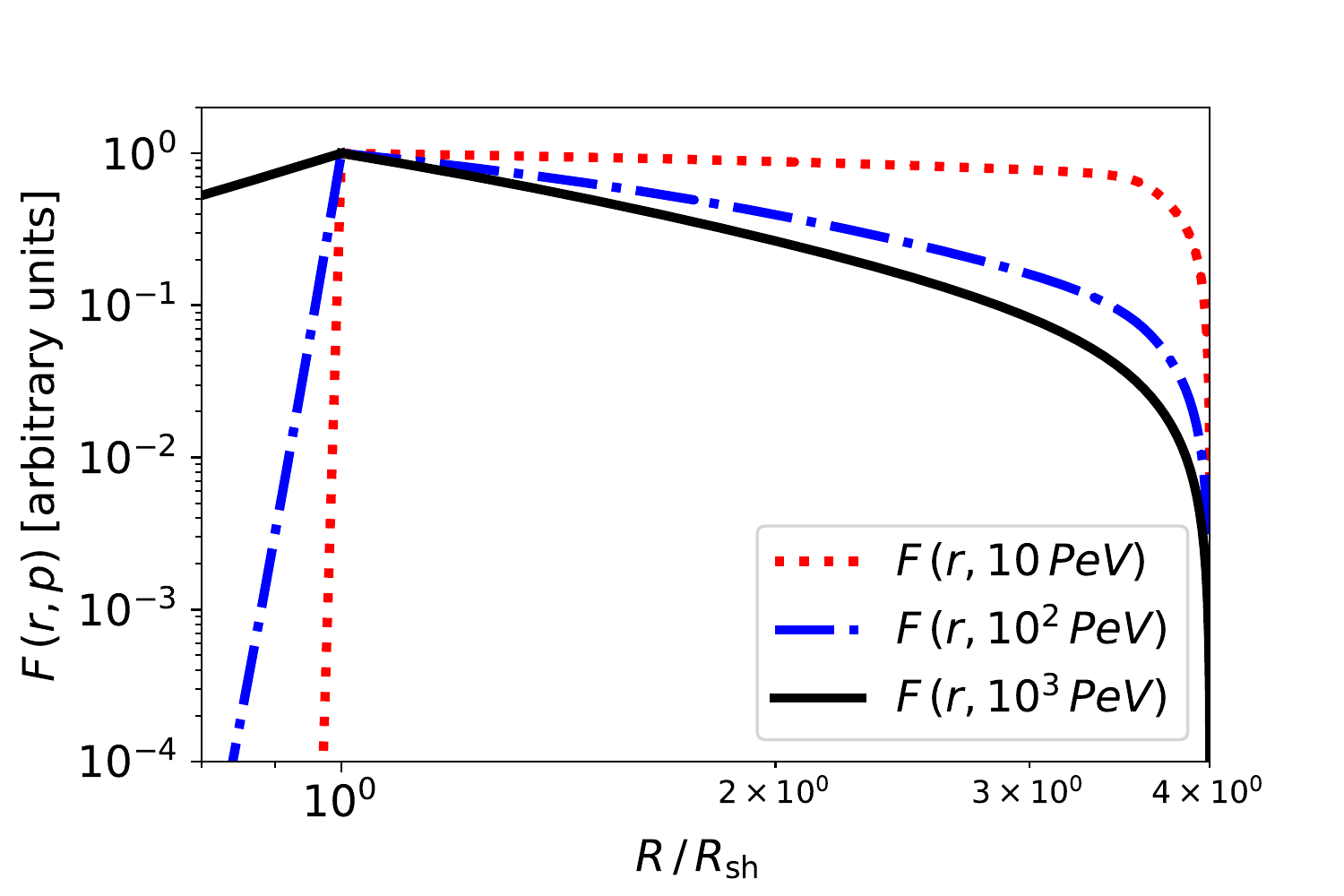}\\
\includegraphics[width=\columnwidth,viewport = -10 0 410 260,clip=true]{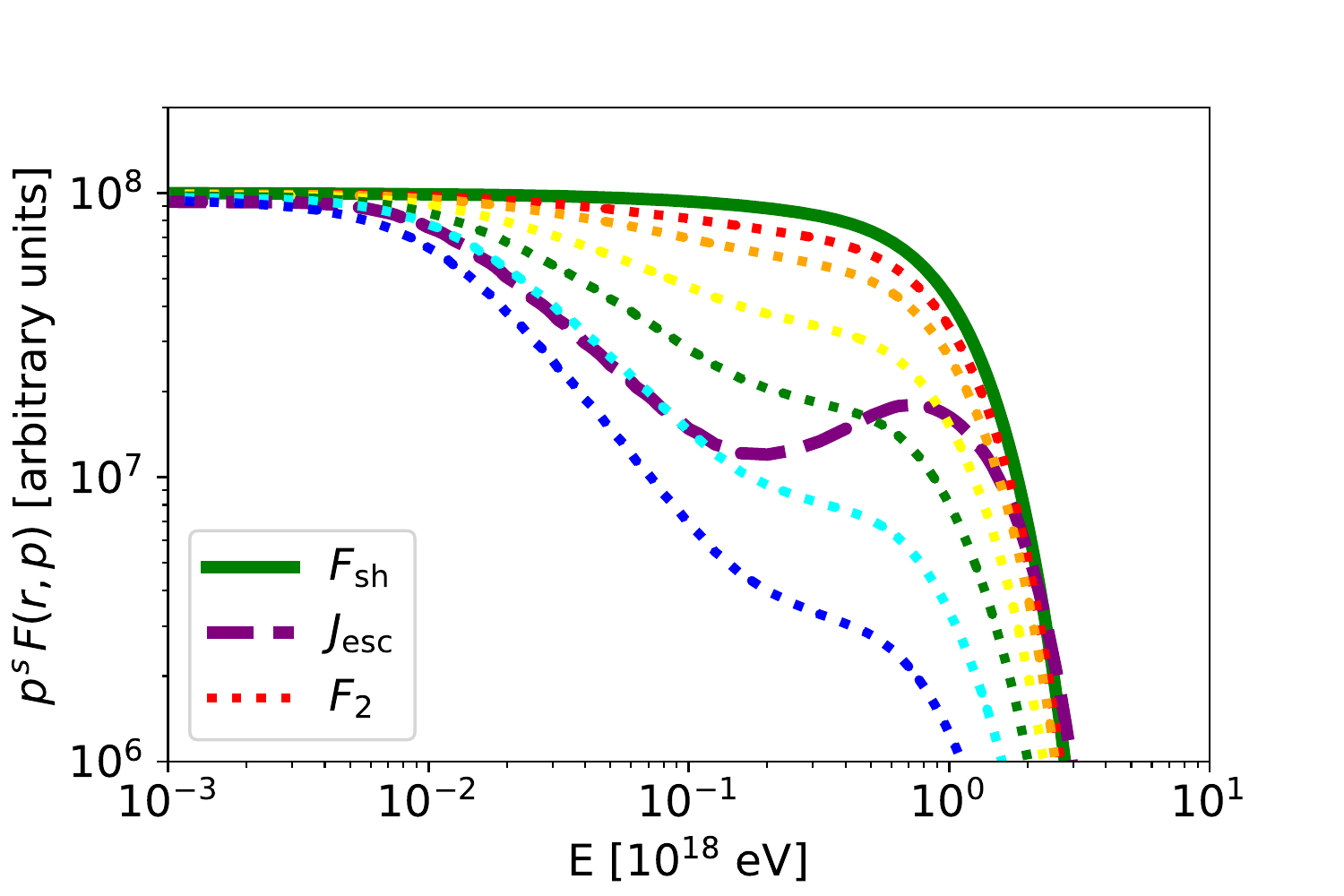}\\[-0.5cm]
\caption[]{{\bf Top panel:} Spatial distribution of the CR phase space density. Low energy particles behave in the system as illustrated by the red dotted line, high energy particle behavior is represented by the blue dot-dashed curve while the black curve shows the behavior of particles at the maximum energy. {\bf Bottom panel:} Spectrum of particles at the shock (thick green line) compared to the spectral shape of the escaping flux (dashed magenta line). The dotted curves represent the particle spectra in the downstream region. From red to blue the dotted lines are computed at $r/R_{\rm fs}=$ 0.29, 0.33, 0.44, 0. 57, 0.75, 0.9.}
\label{fig:Image_2}
\end{figure}

The spatial transport of particles is regulated by advection at low energy and by diffusion at the highest energies while energy losses, both adiabatic (in the upstream region) and inelastic collisions, can affect the normalization and/or introduce spectral features. 
The top panel of Figure~\ref{fig:Image_2} shows the spatial distribution for three different CR energies. The upstream region ($R/R_{\rm sh}<1$) is characterized by the competition between diffusion, which tries to homogenize spatially the particles, and advection which prevents low energy particles to diffuse upwind. In particular, one can see that the higher the energy the stronger the impact of diffusion. The red dotted line illustrates the spatial distribution at low energies while the blue and black lines show results for an intermediate energy and near the exponential cut off, respectively. 
In the downstream region ($R/R_{\rm sh}>1$) one can see that, different from the upstream one, advection-dominated transport leads to a spatially homogenized solution whereas diffusion-dominated transport leads to a number suppression while approaching the free escape boundary. This behavior is a natural result of the spherical geometry of the system.

The bottom panel of Figure~\ref{fig:Image_2} illustrates the spectrum of accelerated particles at the shock (thick green line), at different radii in the downstream region (dotted curves where the red one is the closest to $R_{\rm sh}$ while the blue one approaches $R_{\rm fs}$) and the associated spectrum of the escaping flux (purple dashed line). 
The spectrum of accelerated particles at the wind termination shock, as suggested by Equation~\eqref{Eq: Solution-SHock-Text} and as naturally predicted by DSA in a finite system, is a power-law of index $s$ with maximum energy $E_{\rm max} \simeq 1 \, \rm EeV$ and does not show any relevant additional spectral feature. 
On the other hand, the particle spectrum gradually steepens in the downstream region moving from the wind shock to the forward shock as a result of escape and $p \gamma$ energy losses. In the downstream region, energy losses play a crucial role in shaping the spectrum of particles escaping the system. In particular, the photomeson interactions on the big blue bump occur faster than escape at $\sim 10^{17} \, \rm eV$ as one can also deduce from Figure~\ref{fig:Image_timescales}. 
This results in a dip in the spectrum at such energy, whereas at higher energies the escape is more efficient so that the spectrum hardens at the highest energies.

A comment on the maximum energy is in order: the exponential function regulating the cutoff, $\Gamma_{\rm cut}$, accounts for the geometry of the system and loss mechanisms, so that it cannot be simplified as a ratio $E/E_{\rm max}$. Therefore, here we define $E_{\rm max}$ as the energy where $p^s f_{\rm sh}$ is suppressed by one $e$-fold. In what follows we describe in detail the impact of different realizations of the system to the maximum energy.

\subsection{Impact of parameters on the maximum energy}
\label{SubSec: Param-Scan}

A qualitative estimate of the maximum energy set by the geometry of the system can be obtained by comparing the upstream diffusion length, $D_1/u_1$, with the size of such region, $R_{\rm sh}$ \citep[see also][for additional discussion]{Morlino_starcluster,Peretti_wind}. Since at the highest energies $r_{\rm L}$ is already larger than $l_c$ 
one can write the maximum energy as follows:
\begin{align}
    E_{\rm max} &= q_B \sqrt{\frac{6}{c}} \left[ \frac{\epsilon_B \Dot{M} l_c}{R_{\rm sh}} \right]^{1/2} u_1 \nonumber \\
    & \simeq 1.4 \,  {\rm EeV} \, \left( \frac{\epsilon_B}{0.05} 
     \frac{\Dot{M}}{10^{-1} {\rm M_{\odot} yr^{-1}}} 
     \frac{l_c}{10^{-2} \, {\rm pc}}
     \frac{1 \, \rm pc}{R_{\rm sh}}
     \right)^{1/2}
    \frac{u_1}{0.2 \, c}.
    \label{Eq: EMAX}
\end{align}
As one can see from Equation~\eqref{Eq: EMAX}, the maximum energy for DSA at the wind shock of UFOs turns out to be of the order of EeV for standard values of parameters. 

\begin{table}
\caption{Impact on the maximum energy of a parameter variations. All parameters are set to the benchmark UFO values shown in Table~\ref{table:parameters-0} except for those indicated in the first two columns. The last row shows the result for benchmark values for comparison.}
\label{table:parameters-1}
\centering             
\linespread{1.15}\selectfont
\begin{tabular}{c|c|c}
\hline
\hline
Parameter(s) & Variation(s)& $E_{\rm max}$ [EeV] \\
\hline
{$u_1/c$} & $0.03\,\,/\,\,0.1\,\,/\,\,0.3 $ &  $0.03\,\,/\,\,0.31\,\,/\,\,1.86$   \\ 
$ \Dot{M} [M_{\odot} \, {\rm yr}^{-1}]$ & $10^{-2}\,\,/\,\,1$ &  $0.29\,\,/\,\,2.82$ \\
$\epsilon_{\rm B}$ & $0.01\,\,/\,\,0.1$ & $0.53\,\,/\,\,1.41$  \\
$l_c [{\rm pc}]$ & $3 \cdot 10^{-3}\,\,/\,\,10^{-1}$ & $0.81\,\,/\,\,0.24$  \\
$\delta$ & $5/3$ (Kolmogorov) & $1.02$  \\
$t_{\rm age} [{\rm yr}]$ & $10^2\,\,/\,\,10^4\,\,/\,\,10^5\,\,/\,\,10^6$ & $0.58\,\,/\,\,1.12\,\,/\,\,0.88\,\,/\,\,{0.63}$  \\
$n_{0} [\rm cm^{-3}]$ & $10^3$ {\,/ \, ${10^5}$ \, / \, ${10^6}$} & $1.11$ / ${0.75}$ / ${0.26}$  \\
$U_{\rm rad}$ & none / double & $2.04\,\,/\,\,0.77$  \\
($\Dot{M},u_1$) & pessimistic  / optimistic &  $0.01\,\,/\,\,4.53$ \\
\hline  
\multicolumn{2}{c}{no variations (benchmark)} &  $1.06$ \\
\hline                   
\end{tabular}
\linespread{1.0}\selectfont
\end{table}

Table~\ref{table:parameters-1} highlights the impact of different parametric assumptions on the maximum energy. In particular, we see that, according to Equation~\eqref{Eq: EMAX}, $E_{\rm max}$ scales roughly linearly with $u_1$ and with the square root of $\Dot{M}$ and $\epsilon_B$.
The impact of $l_c$ on $E_{\rm max}$ can be understood as follows: when $l_c \gg 10^{-2} \, \rm pc$, the diffusion coefficient is much larger than the benchmark scenario so that the diffusion length reaches the size of the system at lower energies; when $l_c \ll 10^{-2} \, \rm pc $ the energy at which the diffusion coefficient changes regime (from the standard quasi-linear theory $\sim E^{2-\delta}$ to the small pitch-angle scattering regime $\sim E^2$) shifts to lower energies thereby resulting in a larger value of $D$ at the highest energies. Therefore, since at the highest energies, diffusion dominates, a local maximum  in $E_{\rm max}$ appears for $l_c \simeq 10^{-2} \, \rm pc$. 
Different assumptions on the slope of the turbulence cascade (Kolmogorov-like) have a negligible impact on $E_{\rm max}$ because at the highest energies, where $r_{\rm L} > l_c$, diffusion proceeds in a different regime.

We explored a wide range of possible ages of the system: from $10^2$ yr, where it overcomes all relevant timescales being still consistent with the stationary assumption, up to 1 Myr, where it becomes comparable with the AGN duty cycle. We observe that
the age of the system does not have a strong impact on $E_{\rm max}$ which is affected by less than a factor $2$ for the wide range of alternatives considered. 
Interestingly, we notice that for an age $\gtrsim 10^4 \, \rm yr$, the matter accumulated in the SAM layer becomes calorimetric for low energy particles. This could result in a further hardening of the escaping flux.

The density of the external medium has a direct impact on the size of the system while the effect on the maximum energy is non-trivial. 
For densities $n_{0}\lesssim 10^{5}$, the effect on the maximum energy is moderate and follows approximately Equation\eqref{Eq: EMAX} (see also Equation~\eqref{Eq: wind_shock_rad}). 
On the other hand, higher densities ($n_{0}\gtrsim 10^{5}$) result in a smaller size of the system ($R_{\rm sh} \lesssim 0.4 \, \rm pc$) and greater amount of matter accumulated over the age of the system. 
For such a range of density we observe that the SAM layer becomes calorimetric for low energy particles similar to old UFOs. 
Finally, we observe a deviation from Equation\eqref{Eq: EMAX} when the external density is as large as $n_0= 10^6 \, \rm cm^{-3}$ and the wind shock radius reduces to $R_{\rm sh}\approx 0.2 \, \rm pc$. In fact, for such a reduced size, the $p\gamma$ interactions start to be efficient on the big blue bump photon field of the AGN.

Interestingly, as also highlighted in Figure~\ref{fig:Image_timescales}, different assumptions in the photon field highlight a trend which suggests that the $p \gamma$ interactions on the infrared field of the torus regulate the maximum energy. In particular, $E_{\rm max}$ increases by a factor $ 2$ when the photon field is removed, while it decreases when a stronger photon field is considered. 
This suggests that the infrared field of the torus could play a crucial role in regulating the maximum energy achievable in UFOs. 
We finally explore the combined effect of maximum (minimum) values of $u_1$ and $\Dot{M}$ corresponding to a plausible \textit{optimistic} (\textit{pessimistic}) scenario. 
In this context one can see that UFOs can be responsible for proton acceleration with $E_{\rm max}$ ranging from 10 PeV up to 5 EeV. 
In particular, the objects in the high luminosity end of a hypothetical luminosity function of UFOs are candidate acceleration sites of UHECRs where protons could reach a few EeV. 
Heavier nuclei could be accelerated to higher total energies provided they survive photodisintegration.
The latter possibility depends on the photon background present at the acceleration site and on the relative distance between $R_{\rm sh}$ and $R_{\rm fs}$. 

Our benchmark scenario as well as the scenarios characterized by larger size (resulting from smaller $n_0$ or larger $t_{\rm age}$) are not affected by our approximate treatment of the spatial behavior of the torus photon field because $R_{\rm sh} \gtrsim \rm 1 pc$. 
On the other hand, scenarios characterized by high density and/or younger age could, in principle, be partially affected since $R_{\rm sh}$ would be located at distances smaller than the typical size of the dusty torus. 
Nevertheless, we do not expect a strong impact of different radial dependence of the torus photon field for the following reasons: 1) HE particles cannot access the innermost upstream region due to the dominant effect of advection; 2) since it also scales as $r^{-2}$, the optical big blue bump becomes the dominant thermal photon field for energy losses and limits the maximum energy to be lower than $300$ PeV for $R_{\rm sh}\lesssim 0.2 \, \rm pc$. 
Finally, the gamma-ray absorption is moderately affected by different assumptions on the torus photon field because sub-TeV photons are absorbed on the optical-UV field while beyond 10 TeV they are absorbed by the EBL \citep{Franceschini-EBL} en route to Earth.

\section{Gamma-rays and HE neutrinos from UFOs and constraints to their local density}
\label{Sec: Gamma and Nu}

\begin{figure}
	\includegraphics[width=\columnwidth,viewport=0 0 1100 750,clip=true]{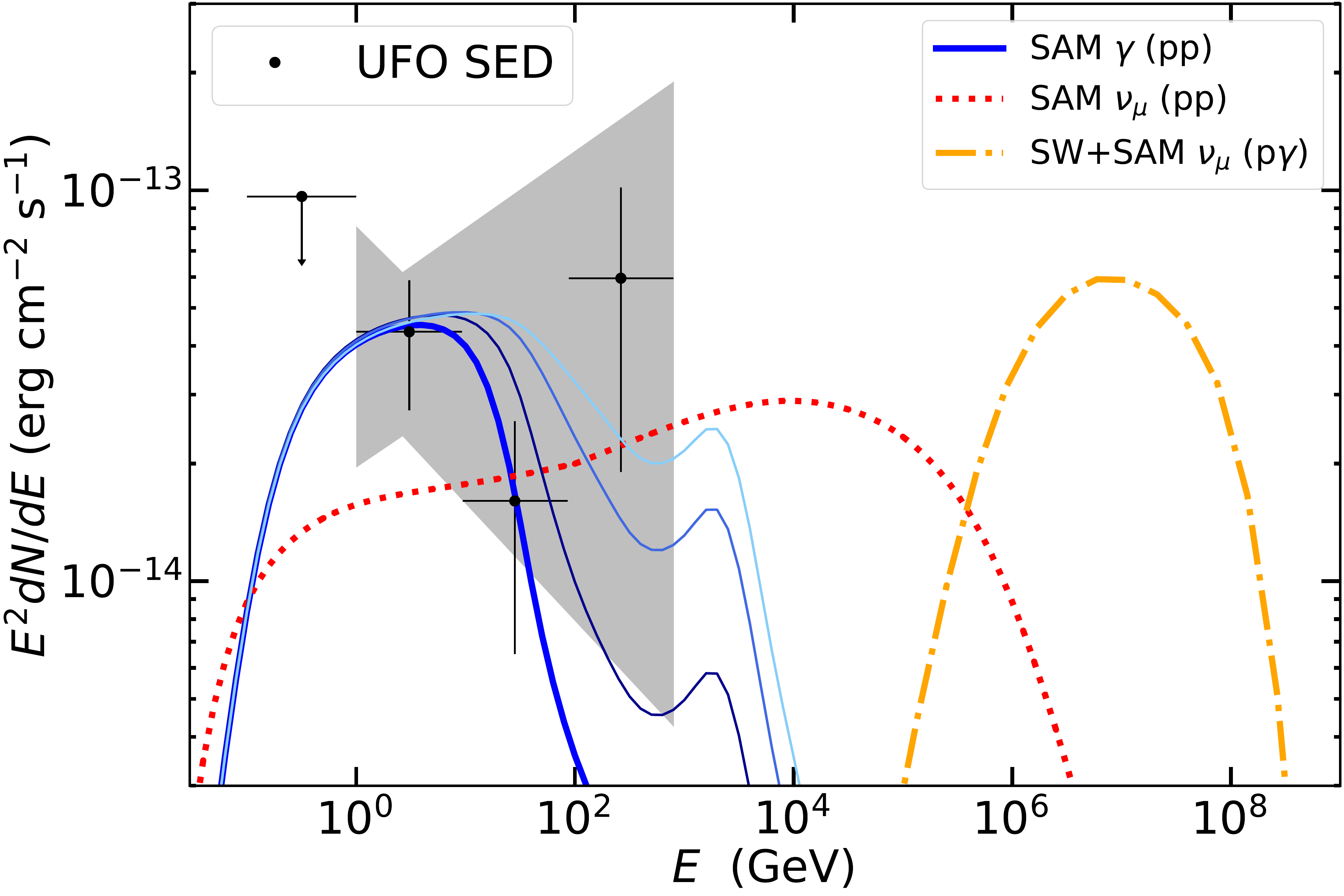}\\[-0.5cm]
	\caption[]{Gamma-ray (thick blue line) and HE neutrino flux produced in the benchmark UFO ($R_{\rm sh}\simeq 0.8 \, \rm pc$) via pp (dotted red line) and p$\gamma$ (orange dot-dashed line) interactions. 
    The acronym SAM (SW) refers to the shocked ambient medium (shocked wind).
	The thin blue-to-cyan lines represents the gamma-ray flux computed respectively for scenario A, B and C (see Table~\ref{table:parameters-0}) in order to illustrate the dependence of the gamma-ray absorption on the bubble expansion (where $R_{\rm sh} \simeq 2, \, 5 \, {\rm and} \, 8  \, \rm pc$ respectively).
	The UFO is assumed to be located at z=0.013 in order to be directly compared with the best-fit UFO SED provided in \citet{UFO-Fermi-LAT+Caprioli}, where the gray band represent the 1 $\sigma$ uncertainty band of such a best-fit UFO SED. }
    \label{fig: Gamma and neutrinos}
\end{figure}

The gas swept-up from the dense environment of the SMBH as well as the strong radiation field of the AGN can make hadronic interactions observationally relevant in UFOs. 
Since interactions are copiously taking place in the UFO wind bubble, a high level of gamma-ray and HE neutrino emission can be expected. 
Figure~\ref{fig: Gamma and neutrinos} illustrates the gamma-ray (thick blue line) and the single-flavor neutrino flux (dotted red for pp and dot-dashed orange for p$\gamma$) flux expected from the benchmark UFO placed at a redshift z=0.013. In particular, the gamma-ray flux is compared with the typical UFO spectral energy distribution (SED) as found in \citet{UFO-Fermi-LAT+Caprioli}.

Despite the fact that the benchmark scenario represents an average UFO in terms of power and maximum energy, the gamma-ray flux of the benchmark UFO ($R_{\rm sh} \simeq 0.8 \, \rm pc$ and $R_{\rm fs} \simeq 3 \, \rm pc$) cannot be representative of the whole class due to the strong radial-dependence of the $\gamma \gamma$ absorption. 
Therefore, in Figure~\ref{fig: Gamma and neutrinos} we also compare it with the expected gamma-ray fluxes as predicted from scenario A ($R_{\rm sh} \simeq 2 \, \rm pc$ and $R_{\rm fs} \simeq 8 \, \rm pc$), B ($R_{\rm sh} \simeq 5 \, \rm pc$ and $R_{\rm fs} \simeq 22 \, \rm pc$) and C ($R_{\rm sh} \simeq 8 \, \rm pc$ and $R_{\rm fs} \simeq 40 \, \rm pc$) as described in Table~\ref{table:parameters-0}. Scenarios A, B and C do not differ from the benchmark scenario in terms of total power but illustrate alternative realizations of it, having a larger size resulting from a longer evolution in a less dense environment. One can observe that these scenarios enhance the gamma-ray emission above $\sim 10$ GeV due to a weaker gamma-gamma absorption and allow a better agreement with the UFO sample observed by \citet{UFO-Fermi-LAT+Caprioli}.

Regardless of the age of the system, gamma-rays of energy greater than a few TeV are completely absorbed by the infrared radiation field of the torus and on the EBL. 
Therefore, pp neutrinos in the 10 TeV - 1 PeV band as well as p$\gamma$ neutrinos in the energy band $10^2$ TeV - $10^2$ PeV would be produced without their gamma-ray counterpart. 
UFOs are thus expected to be bright neutrino sources featuring spectra as hard as $\sim E^{-2}$, while being highly opaque to TeV (and possibly $10-10^2$ GeV) gamma-rays \citep[][]{Susumu_Inoue_NGC1068}.

If the AGN activity during the duty cycle were intermittent, particles leaving the active UFO could end up in a larger scale expanding slower wind, that might result from a previous UFO. 
In this scenario, depending on the local diffusion coefficient experienced by HE particles, adiabatic losses might play a role. 
However, if the local diffusion were comparable to the one in the undisturbed ISM or somehow in between the undisturbed ISM in our Galaxy ($D \approx 10^{28} E_{\rm GeV}^{1/3} \, \rm cm^2 \, s^{-1}$) and the active UFO, the effect on the highest energy particles would be moderate.
On the other hand, if the UFO lifetimes were longer than our assumption for the benchmark scenario the SAM layer could increase luminosity with time \citep[see also][]{Peretti_wind} and eventually turning calorimetric for low energy particles.

UFOs could be common in nearby luminous infrared galaxies (LIRGs) such as active starburst galaxies and Seyfert galaxies. 
However, the abundance and distribution of these objects throughout the Universe as well as their luminosity function are poorly known. 
Therefore, in what follows we estimate the order of magnitude of their diffuse multimessenger emission in terms of EeV cosmic rays and associated HE neutrinos and gamma rays.
Since the horizon for the Bethe-Heitler pair-production suffered by UHECRs on the cosmic microwave background (CMB) is placed beyond $z>2$ we neglect such a loss mechanism in our calculations.

We first assume as a prototype UFO the EeV-atron described by the \textit{benchmark scenario} presented in Table~\ref{table:parameters-0}. As discussed in Appendix~\ref{Appendix: Transport-solution}, the escaping flux is regulated by the interplay between diffusion, advection and energy losses. However, despite its complex analytic expression, assuming an $\sim E^{-2}$ spectrum the power contained by the escaping particles can be approximated as follows:
\begin{align}
\label{Eq: CR Lumin}
    L_{\rm CR} & = \int dp \, 4 \pi p^2 [pc] 4 \pi R_{\rm esc}^2 j_{\rm esc} \simeq \, \frac{3}{4} \, \xi_{\rm CR}  \, \eta_{\rm loss} \, \Dot{M} \, u_1^2 \nonumber \\ 
    & \simeq \, 2 \cdot 10^{43} \,  {\eta_{\rm loss}}  \frac{\xi_{\rm CR}}{0.05}    \frac{\Dot{M}}{0.1 \, \rm M_{\odot}/{\rm yr}}  \left( \frac{u_{1}}{0.2 \, c} \right)^2  \frac{\rm erg}{\rm s}\,,
\end{align}
where $j_{\rm esc}$ is the escaping flux of protons as defined in Appendix~\ref{Appendix: Transport-solution}, $\eta_{\rm loss}\leq1$ is an age-dependent parameter which accounts for the relative reduction in the escaping flux due to energy losses, while the other parameters are normalized to the values shown in Table~\ref{table:parameters-0}. 
The CR luminosity is related to the CR spectral injection rate $\mathcal{Q}_{\rm CR}$ (units of ${\rm GeV}^{-1}{\rm s}^{-1}$) as $L_{\rm CR}=\int {\rm d}E \, E \mathcal{Q}_{\rm CR}(E)$. For simplicity, we assume in the following that the CR emission follows $E^{-2}$ from GeV to EeV, such that $E_p^2 \mathcal{Q}_{\rm CR} \simeq \chi L_{\rm CR}$ with $\chi \equiv 1/\ln({\rm EeV/GeV}) \simeq 0.05$.

In general, the locally observed CR spectrum $\phi_{\rm CR}$ (units of ${\rm GeV}^{-1}{\rm cm}^{-2}{\rm s}^{-1}{\rm sr}^{-1}$) is related to the spectral emission rate of extragalactic sources via a set of transport equations. For CR protons in the EeV energy range we can assume that the transport is dominated by continuous energy loss due to the expansion of the Universe while we neglect the effect of intergalactic magnetic fields. 
Following the notation of \citet{Ahlers2018}, we can estimate the local contribution of UFO EeV-atrons as:
\begin{equation}
\label{Eq: Ahlers_Halzen2}
  [E^2_p\phi_{\rm CR}]_{{\rm EeV}}\simeq\frac{\xi_z}{4 \pi} \frac{c}{H_0} \rho_0  [E^2_{p}\mathcal{Q}_{\rm CR}]_{{\rm EeV}}\,.
\end{equation}
The factor $\xi_z$ is of order unity accounting for the integral in redshift of the source distribution. 
In particular, $\xi_z \simeq 0.5$ (2.6,7.8) for a flat (star-formation rate, AGN) distribution, where the factor for the AGN distribution has been computed for ${\rm Log}L_X=44-45$ adopting data from \citet{Ueda_X_AGN}.  
The parameter $\rho_0$ represents the local comoving density of sources for which we assume $\rho_0 = 10^{-5} \, \rm Mpc^{-3}$ as a reference. Such a value is often quoted as a typical density of AGN with X-ray luminosity of the order of $L_X \simeq 10^{44} \, \rm erg \, s^{-1}$ \citep[see e.g.][]{Fiore_AGN,Ueda_X_AGN,Murase_Waxman_2016}. It is worth mentioning that a source density $\sim 10^{-4} - 10^{-5} \, \rm Mpc^{-3}$ matches also the number density inferred for powerful starbursts such as luminous and ultra-luminous infrared galaxies \citep[see][]{Gruppioni_LF,Peretti2020,Condorelli2022} which are currently also considered to be plausible hosts for UHECR accelerators \citep[][]{Auger_starburst}. 
Expressing the contribution of EeV CR protons to the spectral emission as $[E_p^2 \mathcal{Q}_{\rm CR}]_{{\rm EeV}} = \chi L_{\rm CR}$ with $\chi \simeq 0.05$ we arrive at:
\begin{align} 
     [E^2_p\phi_{\rm CR}]_{{\rm EeV}} \simeq 3\cdot10^{-7} \, \frac{\rho_{0}}{10^{-5} \, \rm Mpc^{-3}} \, \frac{\xi_{z}}{2.6} \, \frac{L_{\rm CR}}{10^{43} \rm erg \, s^{-1}} \,\rm \frac{GeV}{cm^2 s \, sr}\,.
\end{align}
In comparison, the observed CR spectrum has value of $[E^2\phi_{\rm CR}]_{{\rm EeV}} \simeq 2\cdot10^{-7}{\rm GeV}{\rm cm}^{-2}{\rm s}^{-1}{\rm sr}^{-1}$, very close to our estimate.

The associated all-flavor neutrino spectral injection rate $\mathcal{Q}_{\nu}$ resulting from pp interactions can be related to $\mathcal{Q}_{\rm CR}$ as:
\begin{align}
\label{Eq: Nu_mu Lumin}
     E^2_\nu \mathcal{Q}_{\rm all\,\nu}&\simeq \frac{1}{2}  \kappa \, \sigma_{\rm pp} \, c \,  t_{\rm age}\, n_{\rm ISM} \, \eta_{\rm loss}^{-1} \, E_p^2 \mathcal{Q}_{\rm CR}\nonumber \\
    &\simeq 0.07 \frac{n_{\rm ISM,b}}{10^{4} \rm cm^{-3} } \frac{t_{\rm age}}{10^{3} \rm yr} \, \eta_{\rm loss}^{-1} \, E_p^2 \mathcal{Q}_{\rm CR}\,,
\end{align}
where $\kappa\simeq 0.5$ is the inelasticity of pp interactions with cross section $\sigma_{\rm pp}\simeq 3\cdot 10^{-26}{\rm cm}^2$ and the CR proton energy is related to neutrino energy as $E_\nu \simeq 0.05E_{\rm p}$. 
Notice that in this order of magnitude estimate we consider only neutrinos resulting from pp interaction since they dominate up to at least $10^2 \, \rm TeV$. Neutrinos from p$\gamma$ as previously shown, do not feature a different order of magnitude in the HE SED therefore the pp neutrino flux can be assumed as a good approximation also for the order of magnitude flux reached by the p$\gamma$ component.

Again, following \citet{Ahlers2018}, we can now estimate the isotropic neutrino flux as:
\begin{equation}
\label{Eq: Ahlers_Halzen}
  E^2_\nu\phi_{\rm all \nu}= \frac{\xi_z}{4 \pi} \frac{c}{H_0} \rho_0  E^2_\nu\mathcal{Q}_{\rm all \nu}\,.
\end{equation}
Adopting in Equation~\eqref{Eq: Ahlers_Halzen} the expressions for the CR and neutrino luminosity as described in Equation~\eqref{Eq: CR Lumin} and Equation~\eqref{Eq: Nu_mu Lumin} together with $E_p^2 \mathcal{Q}_{\rm CR} \simeq \chi L_{\rm CR}$ one obtains an estimate of the total single flavor neutrino flux at Earth:
\begin{multline} 
     E^2_\nu\phi_{\rm all \nu} \simeq 2\cdot10^{-8} \frac{\rm GeV}{\rm cm^2 s \, sr}\\\times  \frac{\rho_{0}}{10^{-5}  \rm Mpc^{-3}}  \frac{\xi_{z}}{2.6} \frac{n_{\rm ISM,b}}{10^{4} \rm cm^{-3} } \frac{t_{\rm age}}{10^{3} \rm yr} \frac{L_{\rm CR}}{10^{43} \rm erg \, s^{-1}}\,.
\end{multline}
This flux is comparable to the level of the diffuse neutrino flux observed by IceCube \citep[see][]{IceCube_NUMU_LAST} for energies larger than 100 TeV.

These estimates indicate that if UFOs were as abundant as typical non-jetted AGN and powerful starbursts they could potentially be the dominant sources of cosmic rays at EeV while also strongly contributing to the diffuse neutrino flux observed by IceCube above 100 TeV. In this context, the associated gamma-ray flux would not be expected to contribute substantially to the diffuse gamma-ray flux observed by Fermi-LAT \citep[][]{Fermi-LAT-diffused} due to the flat spectral shape and the strong absorption taking place inside the source environment above $\sim 10^2$ GeV that would limit the energy budget of the electromagnetic cascade in the propagation of the gamma rays to the Earth. In a scenario of minimal absorption (scenario C) one can expect at most a gamma-ray flux at $\lesssim 1$ TeV at the same level of the neutrino one, namely $E_{\gamma}^2 \phi_{\gamma} (E_{\gamma}) \sim E_{\nu_{\mu}}^2 \phi_{\nu_{\mu}} (E_{\nu_{\mu}})$. 

Directly observed UFOs as well as X-ray AGN and luminous infrared galaxies where a UFO can be obscured, represent a promising source class where UHECRs could be accelerated. The detection of a flat ($\sim E^{-2}$) neutrino spectrum extending in the PeV range from the innermost core of these galaxies could serve as a hint to discriminate whether DSA is taking place as we propose in this work.

\begin{figure}
	\includegraphics[width=\columnwidth]{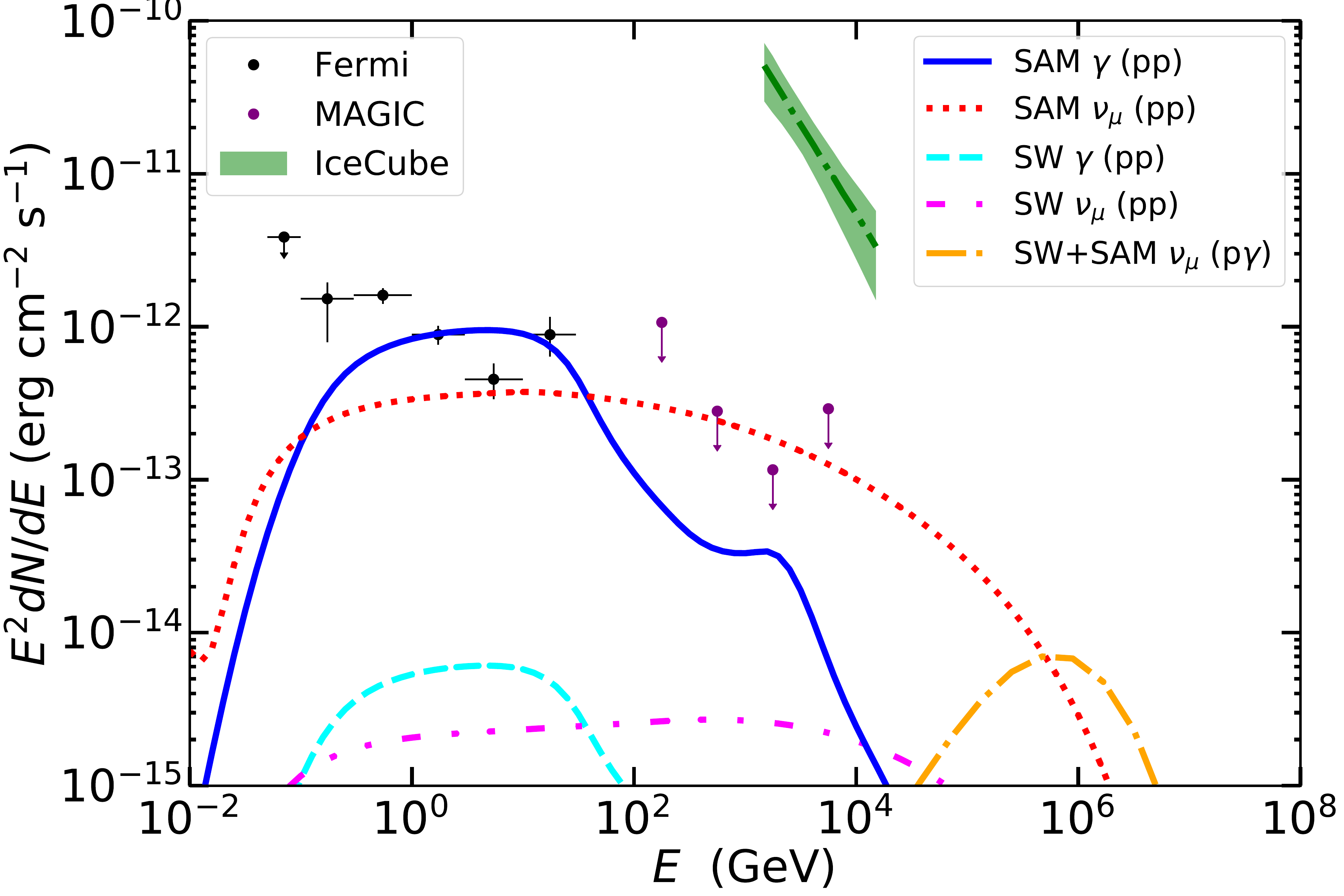}\\[-0.5cm]
	\caption[]{Multimessenger emission for the case of NGC1068. The thick blue line represents the gamma-ray emission dominated by pp interactions in the shocked ambient medium (SAM), while the red dotted line is the associated neutrino flux.
	The emission from the shocked wind (SW) (gamma-ray in cyan dashed and neutrino in magenta dot-dot-dashed) as well as the photomeson (orange dot-dashed) emission are subdominant. The model prediction is compared with Fermi-LAT \citep{Fermi-LAT-CAT}, MAGIC \citep{MAGIC-UL-NGC1068} and IceCube \citep{IceCube-NGC-last} data.}
    \label{fig:Image_3}
\end{figure}

\section{Application to NGC1068}
\label{Sec: Discussion-NGC1068}

In what follows we specialize our calculations to the nearby Seyfert galaxy NGC1068 located at a distance $D_L = 14 \, \rm Mpc$.
Notice that NGC1068 is Compton thick AGN \citep{Matt97}. Therefore, a clear detection of an UFO in its nuclear region is extremely challenging, even though
\citet{Mizumoto2019-NGC1068UFO} reported some indications in that sense.  
Despite one cannot have a compelling evidence of an UFO in NGC1068, in the following we explore the possibility that this galaxy hosts an obscured UFO in its nuclear region.
We compute the gamma-ray and associated neutrino emission through pp and p$\gamma$ interactions by adopting $\Dot{M} = 2 \cdot 10^{-1} \, \rm M_{\odot} \, yr^{-1}$, $u_1 = 0.1 \, c$ and $l_c =10 \, \rm pc$
while assuming all other parameters as in the benchmark scenario. In particular, the coherence length assumed here to be as large as the system size ($2 \, R_{\rm fs} \simeq l_c \simeq 10 \, \rm pc$) results in a maximum energy of approximately $E_{\rm max} = 5 \, \rm PeV$. This is found in agreement with the trend presented in the previous section (\S~\ref{SubSec: Param-Scan}).
Notice that the value of $l_c$ required for this calculation is considerably larger than in our benchmark scenario discussed in Sec.~\ref{Sec: 3-Results}.
In fact, assuming smaller values of $l_c$ leads to a neutrino spectrum extending up to several PeV with an approximately $\sim E^{-2}$ spectrum, in contradiction with the IceCube sensitivity.

Figure~\ref{fig:Image_3} illustrates the multimessenger flux produced by the accelerated particles in the system and under the assumption that the pressure of accelerated particles is $\sim 5 \%$ of the ram pressure at the wind termination shock. One can see that the gamma-ray emission (thick lines) is dominated by the pp contribution in the shocked ambient medium (blue line), whereas the emission from the shocked wind (cyan line) is more than two order of magnitude dimmer. 
As discussed in Sec. \S~\ref{Sec: Gamma and Nu}, the strong photon field associated with the accretion disk and torus makes the source opaque to gamma rays above a few tens GeV, so that all TeV photons are completely absorbed. The neutrino flux is dominated by the pp channel in the range from GeV up to $\sim 10^2$ TeV where it features a reduction due to the maximum energy. 
Photomeson interactions take also place in the UFO environment. However, they produce a negligible impact on the spectrum. Overall the neutrino flux shows a remarkable flat spectrum over more than five orders of magnitude where the associated gamma-ray counterpart gets absorbed beyond $\sim 10 \, \rm GeV$. It is possible to notice that {\it 1)} a UFO could dominate the gamma-ray flux observed by Fermi-LAT and {\it 2)} in the TeV range the UFO could contribute from a few up to $\sim 10\%$ of the flux measured by IceCube~\citep[][]{IceCube-NGC-last} leaving room for other possible sources such as the AGN corona, the molecular outflow or the starburst ring. 
A standard UFO seems therefore disfavoured for explaining the level of neutrino flux observed by IceCube in light of the stringent upper limits imposed by MAGIC as well as the detected Fermi-LAT flux at lower energies.

\subsection{Forward shock scenario for NGC1068}

As an alternative scenario to the acceleration at the wind termination shock one could explore the same UFO at an early stage during which the forward shock is expected to play the most relevant role in terms of particle acceleration.

The forward shock expanding in the unperturbed external medium of density $n_{0} \simeq 10^{4}$ cm$^{-3}$ can indeed foster particle acceleration through DSA. Let us consider a young UFO which has not entered the deceleration phase yet, namely it would be expanding with constant velocity $\sim u_1$. In this context, the forward shock is expected to be strong (${\cal M}\gg 1$) thereby efficient in accelerating particles. The spectrum of accelerated particles at such shock can be written as $f(p) = A (p/p_0)^{-\alpha} \exp (-p/p_{\rm max})$,  where the normalization $A$ is estimated assuming that a fraction $\xi_{\rm fs} \sim 0.1$ of the ram pressure $m_{p}n_{0} u_1^2$ is converted into energized particles. In the test-particle limit, $\alpha =4$, while an upper limit on the maximum momentum can be estimated with the Hillas criterion, leading to:
\begin{equation}
    p_{\rm max, fs} \simeq \frac{\xi_{\rm fs}}{0.1}\, \frac{R_{\rm fs}}{\text{pc}}\, \frac{u_1}{10^8 \text{cm/s}}\,\frac{B}{\mu\text{G}} 10^3 \rm  \, GeV \, c^{-1}\,.
\end{equation}
An acceleration site with $R_{\rm fs} \sim 10^{-2} \, \rm pc$ and $B \sim 10^2-10^3 \, \mu$G allows for $p_{\rm max} \gtrsim 10^{4}-10^{5}$ GeV c$^{-1}$, in principle sufficient for the pp and p$\gamma$ processes to produce gamma rays and neutrinos in an energy range accessible to current instruments. 

Assuming that the accelerated protons fill at most a volume $V \sim 4 \cdot 10^{-6} ({R_{\rm fs}}/{10^{-2}\text{pc}} )^{3}$ pc$^{3}$, in which the interactions with the gas and the AGN photon field are taking place, the UFO is expected to produce a gamma-ray luminosity (before any absorption effects are taken into account) of the order of $L_{\gamma} (10^3-10^4 \text{GeV})\simeq 10^{-15}-10^{-14}$ erg cm$^{-12}$ s$^{-1}$, and a neutrino luminosity of the order of $L_{\nu} (10^3-10^4 \text{GeV})\simeq 10^{-16}-10^{-15}$ erg cm$^{-2}$ s$^{-1}$, indicating that the emission from the forward shock is expected to be sub-dominant compared to the one from the wind termination shock happening at later times. 

Late stage UFOs, as discussed in \citet{UFO-Fermi-LAT+Caprioli}, could be luminous enough to be detected in the GeV range by space based telescopes such as Fermi-LAT. However, this would require the central engine to be active for a time $t \gg 10^3 \, \rm yr$. 
On the other hand, high acceleration efficiency at the wind termination shock can be expected to be found soon after the deceleration phase has begun. 

\subsection{Discussion on NGC1068}
\label{Subs: 1068 disc}

UFOs are characterized by a prominent gamma-ray (up to $\sim 10-10^2$ GeV) and neutrino emission (typically up to $\sim 10^2$ TeV) resulting from pp interactions. Therefore, in the standard test particle regime considered in this work, the resulting gamma-ray and neutrino spectra will feature roughly the spectral slope ($\sim E^{-2}$) of their parent protons. We observe that from the energetic point of view, the level of neutrino flux observed is compatible with the amount of power that UFOs can supply on average suggesting the AGN as a plausible origin for such an emission. However, the strongest limitation comes from the level of gamma rays measured by Fermi-LAT. In fact, even though the AGN radiation field can absorb efficiently TeV photons, GeV gamma-rays come basically unabsorbed. Indeed, as pointed out by several authors \citep[see e.g.][]{Murase_AGN_cores,Inoue2020,Kheirandish_2021,Susumu_Inoue_NGC1068,Foteini_NGC1068,Murase2022}, such level of neutrinos is not compatible with a pp scenario with a $\sim E^{-2}$ spectrum unless the emission comes from a region optically thick to GeV and sub-GeV gamma rays, namely the nearest neighbourhood of the SMBH or AGN-corona having a size of $\lesssim 10^2 \, R_s$ (where $R_s$ is the Schwarzschild radius).

Even though it is not clear whether the launching radius of a UFO can be localized at such a small distance from the SMBH, we explored under which conditions one could expect a UFO to be confined close to the AGN corona for a sufficient amount of time during its deceleration phase. We found that, in order to reproduce the level of neutrino flux inferred by IceCube~\citep[][]{IceCube-NGC-last} without exceeding the gamma-ray flux, the energy budget would not exceed standard values typical of UFOs ($\Dot{E}\lesssim 10^{45} \, \rm erg \, s^{-1}$) but the requirement in terms of average gas density of the external medium would be $n_{\rm ISM} \gtrsim 10^{10} \, \rm cm^{-3}$. 
Such a density could be compatible with the high column density inferred for this source\citep[see e.g.][]{Matt97}, therefore this could be a plausible scenario to account for the observed neutrino flux. 
A detailed modeling of a confined wind expanding in the ultra-dense medium goes beyond the scope of this work, therefore it is left for a follow-up investigation. 

The recent anisotropy study carried out by the Pierre Auger Observatory suggests that Seyfert and Starburst galaxies (or objects with a spatial distribution related to these sources) may be related to this anisotropy \citep[][]{Auger_starburst}. It is intriguing that NGC1068 is the fourth most relevant object in such catalogs, providing a contribution of about $5-10\%$. However, our calculations suggests that if the gamma-ray flux detected by Fermi-LAT is associated with the UFO activity in this source, then the maximum energy is bound to be too low to be relevant for UHECRs. Viceversa, if UHECRs are to be produced in the UFO, a different source for the gamma rays has to be found.

\section{Conclusions}
\label{Sec: 5-Conclusions}

In this work we investigated the potential of diffusive shock acceleration at the wind termination shocks of ultra-fast outflows (UFOs) in the core of active galaxies. We developed a model of acceleration and transport of particles in the wind bubbles excavated by UFOs and we studied the multimessenger implications in terms of escaping particles and high-energy photons and neutrinos produced through hadronic interactions. 

We found that protons can be accelerated up to the EeV range and that the far infrared photon field of the torus could play a dominant role in setting the maximum energy in such sources. 
In addition, the transport condition in the UFO environment could result in a spectral hardening of the escaping flux of the cosmic rays at the highest energies.
Such energetic and spectral properties are crucial in light of the recent results obtained by the Pierre Auger Observatory which suggest that the sources of UHECRs could be characterized by hard spectra \citep[see e.g.][and references therein]{Auger_combined_fit}. 

UFOs are extremely interesting in terms of multimessenger emission since, in standard conditions, they are expected to shine in gamma rays up to $\sim 10$ GeV while being opaque beyond $10-10^2$ GeV depending on the parametric configuration.
While gamma rays are efficiently absorbed, HE neutrinos from pp and p$\gamma$ interactions will be copiously produced with a spectrum extending up to $\sim 10^2$ PeV featuring a spectral slope as hard as the one of their parent protons. 
Such a property is particularly interesting since UFOs have the potential and possibly the number density in the Universe to simultaneously dominate the CR spectrum at the ankle and the diffuse neutrino flux observed beyond $\sim 10^2$ TeV.

We finally applied our model to the UFO that could be present in the nuclear region of NGC1068 and we found out that, if confirmed, in favourable parametric conditions, a well developed spherical UFO wind bubble could dominate the gamma-ray flux observed by Fermi-LAT while it is unlikely to contribute more than $\sim 10 \%$ of the total neutrino flux observed by IceCube at TeV.
On the other hand the base of the outflow could be a plausible region where protons can be injected and produce HE neutrinos while the associated GeV gamma-ray counterpart can be efficiently reprocessed to lower energies \citep[see also][]{Susumu_Inoue_NGC1068}.

The present model focused on the acceleration and transport of protons in UFOs while the study of potential implications for primary and secondary electrons as well as electron-positron pairs is left for future work. UFOs are in fact expected to be perfect electron calorimeters given the dominant synchrotron and inverse Compton timescales compared to the escape timescale, and the case of leptons would require a specific treatment. We computed the typical timescales for electrons and we found that primary electrons cool very rapidly without reaching TeV energies. Secondaries as well as electron-positron pairs are also expected to cool mostly via synchrotron losses since beyond $\sim 1$ TeV the interaction with the big blue bump takes place in the Klein-Nishina regime. 
In such energy range, the electromagnetic cascade is expected to be synchrotron-dominated for standard parametric assumptions ($\epsilon_{\rm B} \gtrsim 10^{-2}$). Therefore, one could expect the leptonic emission to produce some possible signatures in the hard X-ray to soft gamma-ray energy band and in radio.

Heavy nuclei should also be implemented in a future work since their transport, similar to electrons, requires the inclusion of calorimetric conditions, continuous energy losses as well as fragmentation and re-injection at lower masses. While our focus has been on the proton population, it is worth mentioning that nuclei can be expected to be co-accelerated in UFOs with a rigidity dependence in the maximum energy and possibly a higher efficiency in the injection of energetic heavier elements \citep[see e.g.][]{Caprioli-Chemical-DSA}. 
In particular, while protons only partially suffer energy losses, heavy nuclei of electric charge $Z$ are expected to efficiently fragment on the AGN photon field. 
Therefore one can expect that at early time the majority of heavy nuclei would be reprocessed while at later time they could start escaping efficiently with an energy as large as $\sim Z \cdot \rm EeV$.

With the present work we propose UFOs as candidate sources of UHECRs and efficient high-energy neutrino emitters possibly opaque to gamma rays beyond a few tens GeV. Such properties make UFOs a remarkable source class not only for the UHECRs detected by the Pierre Auger Observatory but also for the diffuse HE gamma-ray and neutrino fluxes observed respectively by Fermi-LAT and IceCube. 

\section*{Acknowledgements}

The research activity of EP and MA was supported by Villum Fonden (project No.~18994). EP was also supported by the European Union’s Horizon 2020 research and innovation program under the Marie Sklodowska-Curie grant agreement No. 847523 ‘INTERACTIONS’. FGS acknowledges financial support from the PRIN MIUR project ``ASTRI/CTA Data Challenge'' (PI: P. Caraveo), contract 298/2017. 
GM is partially supported by the INAF Theory grant 2022 ``Star Clusters as cosmic ray factories''. EP is grateful to Antonio Condorelli for insightful discussions.

\section*{Data Availability}
No data has been analyzed or produced in this work.



\bibliographystyle{mnras}
\bibliography{example} 



\appendix

\section{Properties of the AGN wind bubble}
\label{Appendix: Self-similar radii evo}

\begin{figure}
\centering
	\includegraphics[width= \columnwidth,viewport = 0 0 1090 750,clip=true]{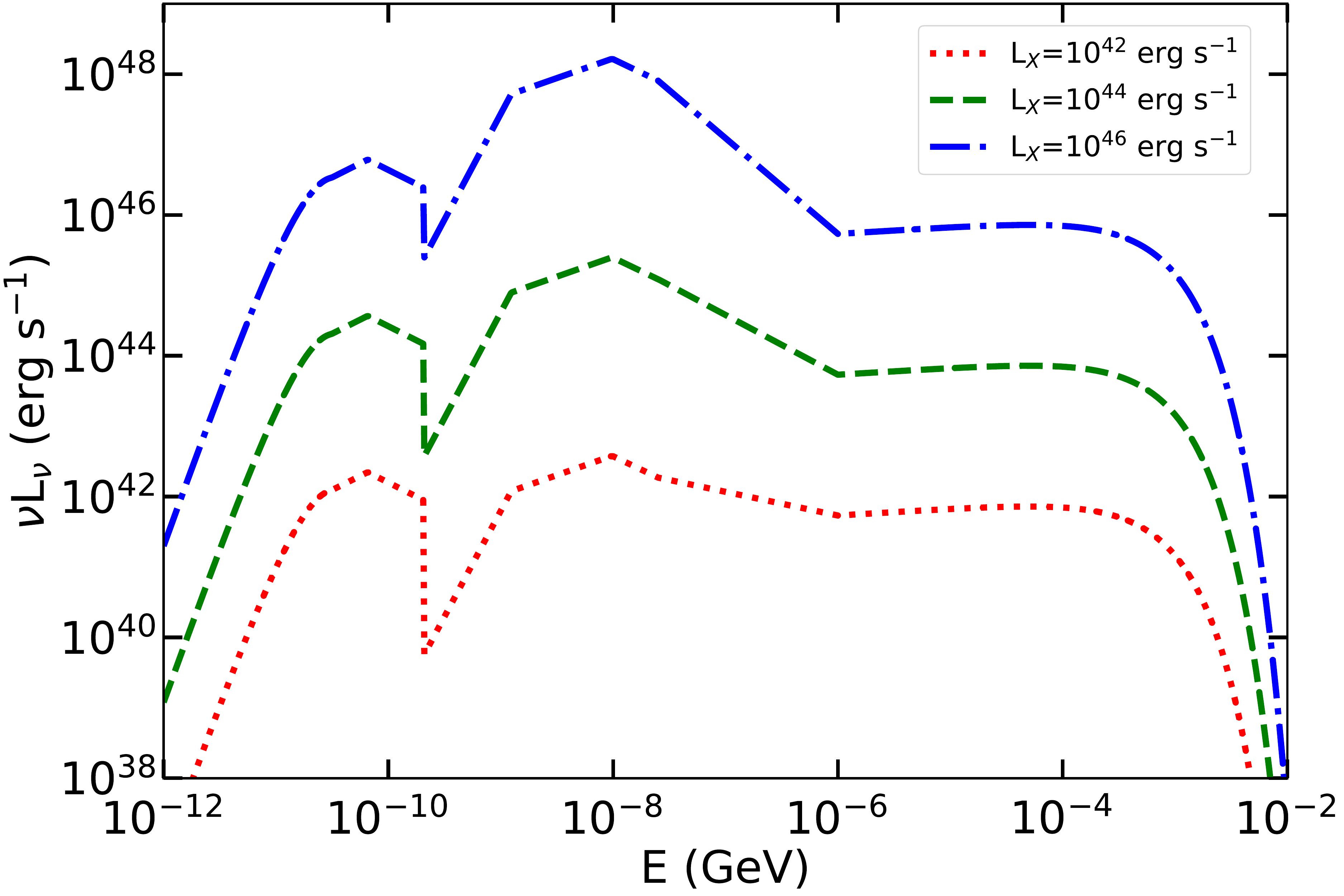}\\\includegraphics[width=\columnwidth,viewport =0 0 1100 750,clip=true]{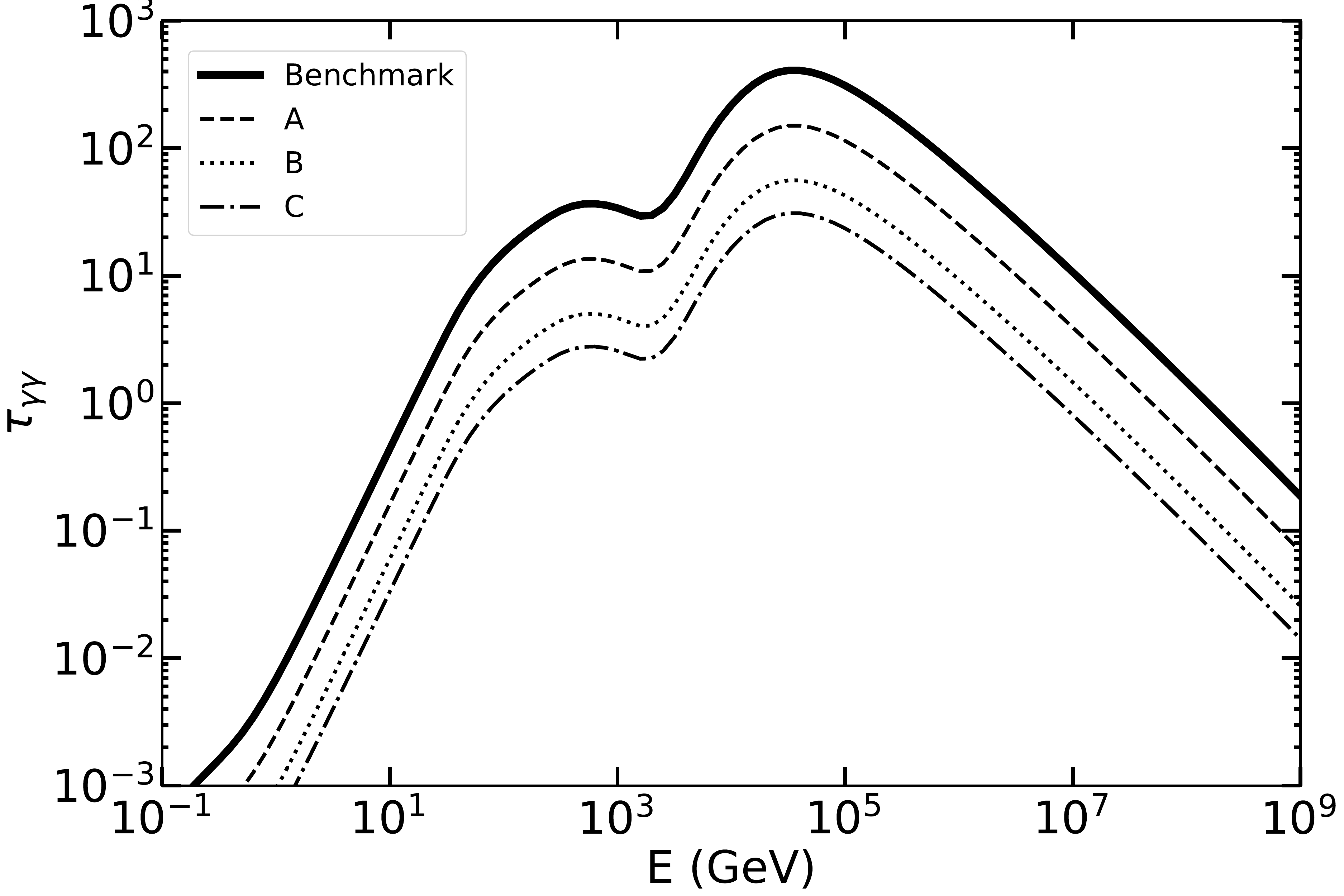}
    \caption[]{{\bf Top Panel:} the AGN photon field for three different X-ray luminosity. Dotted red, dashed green and dot-dashed blue represent the typical spectral energy distribution of an AGN having an X-ray luminosity of $10^{42}$, $10^{44}$ and $10^{46}$ $\rm erg \, s^{-1}$ respectively. {\bf Bottom panel:} gamma-gamma opacity of the wind bubble where the different curves highlight the scenarios discussed in Sec. \S~\ref{Sec: 3-Results}. In particular: the thick black line represents the benchmark scenario while the thin lines describe the absorption in the Scenario A, B and C.}
    \label{fig:Image_BKG_fields}
\end{figure}

The dynamics of the wind shock and forward shock during the deceleration phase of a wind bubble have been studied by \citet[]{Koo-McKee,Koo-McKee2}.
The location of the wind shock is described by the following relation:
\begin{multline}
\label{Eq: wind_shock_rad}
  R_{\rm sh} = 23.0 \, { \rm pc} \, \, \left(\frac{t_{\rm age}}{1{\rm Myr}}\right)^{2/5}\left(\frac{\Dot{E}}{10^{38}{\rm erg}/{\rm s}}\right)^{3/10}  \\\times\left(\frac{n_{0}}{1{\rm cm}^{-3}}\right)^{-3/10} \left(\frac{u_{1}}{10^8{\rm cm}/{\rm s}}\right)^{-1/2},
\end{multline}
where $t_{\rm age}$ is the age of the system, $\Dot{E}$ the kinetic power ($\Dot{E}= \Dot{M}u_1^2 /2 $), $n_0$ the external medium density and $u_{1}$ the upstream wind speed.
The forward shock radius is found as follows:
{\begin{equation}
    R_{\rm fs} = 76.0 \, {\rm pc} \, \, \left(\frac{t_{\rm age}}{1{\rm Myr}}\right)^{3/5} \left(\frac{\Dot{E}}{10^{38}{\rm erg}/{\rm s}}\right)^{1/5}\left(\frac{n_{0}}{1{\rm cm}^{-3}}\right)^{-1/5}.
\end{equation}}
{The radiation field (top panel) and the corresponding opacity (bottom panel) of the wind bubble are shown in Figure~\ref{fig:Image_BKG_fields}. 
In particular, the typical AGN radiation field feautures a combination of thermal and non-thermal components. 
At around $\sim 10 \, \rm meV$ one can identify the infrared thermal field of the torus, while the second thermal component peaking at $\sim 10 \, \rm eV$ is the thermal radiation from the accretion disk known as \textit{big blue bump}. Finally, starting from the keV range the non-thermal tail produced by electrons in the AGN corona is present. Such a non-thermal tail can additionally feature a somehow different shape due to the mechanism known as Compton reflection. In this work we decided to ignore such an effect since it has a strong dependence on the local parameters of the AGN while having a negligible impact on the high energy particles and their radiation.
The gamma-ray opacity shown in the bottom panel clearly highlights the prominent impact of the big blue bump and the torus absorbing respectively in the energy ranges 10 GeV - TeV and beyond 10 TeV. 
}

\section{Solution to the transport equation}
\label{Appendix: Transport-solution}

{The transport equations~\eqref{Eq: Transport} are solved separately in the upstream and the downstream regions and joined at the wind shock location as described in }\citet{Morlino_starcluster,Peretti_wind}. 
In what follows we highlight the main analytical steps required to obtain the formal solution and the key aspects of the iterative algorithm we adopt to find the numerical solution.

\subsection{Upstream region}

By integrating Equation~\eqref{Eq: Transport} from $r=0$ to $r<R_{\rm fs}$ one obtains:
\begin{equation}
    \label{Eq: Up1}
    r^2u_1f_1(r,p) = r^2D_1(r,p) \partial_rf_1(r,p) - G_1(r,p) - H_1 (r,p)\,,
\end{equation}
where the subscript ``1'' refers to the upstream region, while the functions $G_1$ and $H_1$ have the following expressions:
\begin{gather}
    G_1(r,p) = \frac{1}{3} \int_0^r dr' \, \left[- \frac{\partial {\rm ln} [p^3f_1(r',p)]}{\partial {\rm ln} p} \right] \, f_1(r',p) \, \partial_{r'}[r'^2u_1]\,, \\
    H_1(r,p) =  \int_0^r dr' \, \lambda_1(r',p) \, r'^2 \, f_1 (r',p)\,,
\end{gather}
where {$\lambda_1= \tau_{\rm pp}^{-1} + \tau_{{\rm p}\gamma}^{-1}+ \tau_{\rm BH}^{-1}$ 
is the CR energy loss accounting for pion production in p$\gamma$ and pp interactions~\citep[see][]{Kelner_pg} as well as Bethe-Heitler (BH) pair-production~\citep[see, e.g.,][]{Gao2012-BH}.} 
Defining the effective upstream velocity as:
\begin{equation}
    V_{\rm eff,1} (r,p) = u_1 \left[ 1 + \frac{G_1(r,p)+H_1(r,p)}{u_1 r^2 f_1(r,p)} \right],
\end{equation}
the solution of Equation~\eqref{Eq: Up1} is straightforwardly obtained:
\begin{equation}
    \label{Eq: Up-solution}
    f_1(r,p) = f_{\rm sh}(p) {\rm exp} \left[- \int_r^{R_{\rm sh}} dr' \frac{V_{\rm eff,1}(r',p)}{D_1(r',p)} \right]\,.
\end{equation}

\subsection{Downstream region}

Similarly to the upstream case, also in the downstream Equation~\eqref{Eq: Transport} is first approached through a spatial integral exploiting the boundary condition. 
Integrating from
from $r>R_{\rm fs}$ to $r=R_{\rm fs}$ one obtains:
\begin{equation}
    \label{Eq: Down1}
    r^2u_2(r)f_2(r,p) = r^2D_2(r,p) \partial_rf_2(r,p) + R_{\rm fs}^2 j_{\rm esc}(p) + H_2 (r,p)\,,
\end{equation}
where the subscript ``2'' refers to the downstream region, while the function $H_2$ reads:
\begin{equation}
    H_2(r,p) =  \int_r^{R_{\rm fs}} dr' \, \lambda_2(r',p) \, r'^2 \, f_2 (r',p)\,.
\end{equation}
As in the upstream region, it is convenient to define the effective velocity in the downstream region as:
\begin{equation}
    V_{\rm eff,2}(r,p) = u_2(r) \left[ 1 - \frac{H_2(r,p)}{r^2u_2(r)f_2(r,p)} \right]\,.
\end{equation}
It is also useful to define the integral function $I_2$ as:
\begin{equation}
    I_2(r,p) = \int_{R_{\rm sh}}^r \frac{dr'}{r'^2} e^{-\varphi_2(r',p)}\,,
\end{equation}
where $\varphi_2(r,p)$ has the following expression:
\begin{equation}
    \varphi_2(r,p) = \int_{R_{\rm sh}}^{r} dr' \frac{V_{\rm eff,2}(r',p)}{D_2(p)}\,.
\end{equation}
Integrating Equation~\eqref{Eq: Down1}  one obtains the expression for the escaping flux and the downstream solution as:
\begin{gather}
\label{Eq: Down_Solution}
    f_2(r,p) = f_{\rm sh}(p) \, e^{\varphi_2(r,p)} \left[ 1 - \frac{I_2(r,p)}{I_2(R_{\rm fs},p)} \right] \\
    j_{\rm esc}(p) = f_{\rm sh}(p) \frac{D_2(p)}{R_{\rm fs}^2 I_2(R_{\rm fs},p)} \,.
\label{Eq: Esc_flux}    
\end{gather}
We finally notice that the escaping flux $j_{\rm esc}$ can be rewritten as 
\begin{equation}
    j_{\rm esc}(p) = \eta_{\rm loss} \frac{u_2 f_{\rm sh}(p)}{1-\exp[-R_{\rm sh}u_2(1-R_{\rm sh}/R_{\rm fs})/D_2(p)]} \frac{R_{\rm sh}^2}{R_{\rm fs}^2}
\end{equation}
where $\eta_{\rm loss}$ is a parameter $\lesssim 1$ accounting for energy losses. 

\subsection{Solution at the shock}

The shock solution is obtained by integrating Equation~\eqref{Eq: Transport} across an infinitely small layer embedding the wind shock. 
The result is the following: 
\begin{equation}
\label{Eq: Sh1}
    [D_2 \partial_rf_2 - D_1 \partial_r f_1]_{r=R_{\rm sh}} - \frac{u_1-u_2}{3} p \partial_p f_{\rm sh}(p) + Q_0 (p) = 0\,.
\end{equation}
Substituting Equation~\eqref{Eq: Up1} and Equation~\eqref{Eq: Down1} in the first term on the left-hand side, Equation~\eqref{Eq: Sh1} can be rewritten as:
\begin{align}
\label{Eq: Sh2}
    \frac{s Q_0(p)}{u_1} & = p \partial_p f_{\rm sh}(p) + sf_{\rm sh}(p) + s \Psi_{ l}(p)  f_{\rm sh}(p)
    + s \Psi_{ e}(p) f_{\rm sh}(p)\,,
\end{align}
where the functions $\Psi_k$ ($k=l,e$) are defined as:
\begin{gather}
    \Psi_{ l}(p) = \frac{{G}_1(R_{\rm sh},p)+{H}_1(R_{\rm sh},p)+{H}_2(R_{\rm sh},p)}{u_1 R_{\rm sh}^2 f_{\rm sh}(p)}\,, \\
    \Psi_{ e}(p) = \frac{[D_2 {I}_2^{-1}(R_{\rm sh},p)-R_{\rm sh}^2 u_2]}{u_1 R_{\rm sh}^2}\,.
\end{gather}
Here the subscripts $l$ and $e$ stands for loss and escape respectively.
Finally, by recognizing a total derivative on the right hand side of Equation~\eqref{Eq: Sh2}, the solution at the shock {can be} obtained as:
\begin{equation}
    \label{Eq: Shock_Solution}
    f_{\rm sh}(p) = \frac{s \eta n_1}{4 \pi p_{\rm inj}^3} \left( \frac{p_{\rm inj}}{p} \right)^s e^{-\Gamma_l(p)} e^{-\Gamma_e(p)}\,,
\end{equation}
where:
\begin{equation}
    \Gamma_{l(e)}(p) = s \int_{p_{\rm inj}}^p \frac{dp'}{p'} \Psi_{l(e)}(p')\,.
\end{equation}
{Notice that Equation~\eqref{Eq: Shock_Solution} can be re-written in the compact form 
(see Equation~\eqref{Eq: Solution-SHock-Text}) $f_{\rm sh}(p) = C \, p^{-s} {\rm exp}[-\Gamma_{\rm cut}(p)]$, where $C= s\, \eta\, n_1\, p_{\rm inj}^{s-3}/(4 \pi)$ and $\Gamma_{\rm cut} = \Gamma_l + \Gamma_e$.}

\subsection{Iteration algorithm}

The solution to the transport equation on the two sides of the shock, $f_1$ (Equation~\eqref{Eq: Up-solution}) and $f_2$ (Equation~\eqref{Eq: Down_Solution}), and at the shock, $f_{\rm sh}$ (Equation~\eqref{Eq: Shock_Solution}), do not have a simple analytic form since they depend on each other through the functions $V_{\rm eff,1}$, $V_{\rm eff,2}$ and $\Psi_{l(e)}$, respectively. {A solution can be found found via an iterative algorithm.}

{We initialize the solutions for the set of functions ($f_{\rm sh}^{(0)}$,$f_1^{(0)}$,$f_2^{(0)}$) by the solutions resulting from the}following no-loss conditions: $G_1^{(0)} = H_1^{(0)} = H_2^{(0)} = 0$; $V_{\rm eff,1}^{(0)}(r,p)= u_1$; $V_{\rm eff,2}^{(0)}(r,p)= u_2(r)$.
This results in $\Psi_l^{(0)} = 0$ while  $\Psi_e^{(0)}$ reduces to the following analytic form:
\begin{equation}
    \Psi_e^{(0)}(p) = \frac{u_2/u_1}{ {\rm exp}\left[\frac{R_{\rm sh}u_2}{D_2(p)}\left(1-\frac{R_{\rm sh}}{R_{\rm fs}}\right)\right] - 1 }\,.
\end{equation}
{We start from this initial approximation and find iterative solutions by re-computing all functions with the set of solution of the previous iteration, namely:}
\begin{align*}
   \left(f_{\rm sh}^{(i)},f_1^{(i)},f_2^{(i)}\right) & \to \left(G_1^{(i+1)}, H_1^{(i+1)}, H_2^{(i+1)}\right) \\
   & \to \left(V_{\rm eff,1}^{(i+1)}, V_{\rm eff,2}^{(i+1)}, \Psi_l^{(i+1)},\Psi_e^{(i+1)}\right)  \\ & \to
   f_{\rm sh}^{(i+1)} \\ & \to \left(f_1^{(i+1)},f_2^{(i+1)} \right)
\end{align*}
where $(i)$ and $(i+1)$ indicate the i-th and (i+1)-th iteration. 
This algorithm is repeated until the phase space density at the n-th iteration $f^{(n)}$ is indistinguishable from the solution found at the iteration (n-1)-th, $f^{(n-1)}$, namely when a convergence condition has been obtained.


\bsp	
\label{lastpage}
\end{document}